\documentclass[arxiv,a4paper,fleqn]{cas-dc}

\usepackage[authoryear]{natbib}
\usepackage{url}
\usepackage{booktabs}
\usepackage{threeparttable}
\usepackage{graphicx}
\usepackage{lineno}

\def\tsc#1{\csdef{#1}{\textsc{\lowercase{#1}}\xspace}}
\tsc{WGM}
\tsc{QE}

\newcommand{\pv}{$p_V$}

\newcommand{\sigerr}{$\sigma_{\rm err}$}
\newcommand{\sigobs}{$\sigma_{\rm obs}$}
\newcommand{\sigint}{$\sigma_{\rm int}$}
\newcommand{\neowise}{\textit{NEOWISE}}
\newcommand{\akari}{\textit{AKARI}}
\newcommand{\sdss}{SDSS}
\begin{document}
\let\WriteBookmarks\relax
\def\floatpagepagefraction{1}
\def\textpagefraction{.001}

\shorttitle{Albedo Structure in Asteroid Families}    

\shortauthors{M. Kaplan}  

\title [mode = title]{Statistical Constraints on Albedo Structure in Asteroid Families from NEOWISE Measurements}  



%

\author{Murat Kaplan}[orcid=0000-0003-2595-5463]



\ead{muratkaplan@akdeniz.edu.tr}


\credit{Writing -- original draft, Software, Methodology, Conceptualization}

\affiliation{organization={Akdeniz University},
            addressline={Department of Space Sciences and Technologies}, 
            city={Antalya},
            postcode={07058}, 
            country={Türkiye}}








\begin{abstract}
Interpreting asteroid family albedo distributions as compositional signatures 
requires distinguishing intrinsic structure from measurement artifacts---but 
this is rarely quantified. We analyze 102 families using NEOWISE data with 
AKARI cross-validation and SDSS-based photometric checks, establishing detection 
limits for bimodality and quantifying selection bias in size--albedo correlations.

We work primarily in logarithmic albedo space ($\log_{10} p_V$), which more 
accurately resolves the asymmetric dark/bright mixtures characteristic of compositionally 
mixed families. NEOWISE albedos are error-dominated. Measurement uncertainties 
rival observed spreads (median $\sigma_{\rm obs}/\sigma_{\rm err} \approx 0.99$ 
in log space). AKARI cross-validation on 1{,}498 matched asteroids confirms 
NEOWISE measurements broadly (slope $=1.03$, $R^2 = 0.76$, median 
$|\Delta p_V| = 0.015$), with a known spectral-type-dependent offset of $+11\%$ 
on average. Genuine bimodality meeting conservative criteria is found in 6 of 
102 families (5.9\%); the two most secure detections (Nysa-Polana, Juno) are 
recovered in both linear and logarithmic albedo, the other four only in log.
Monte Carlo simulations show that a 
dark/bright bimodality at the characteristic separation ($\Delta\mu \approx 
0.64$ dex) is recovered with high probability for the intrinsic scatter present 
in the families, so the low rate reflects genuine compositional homogeneity 
rather than limited sensitivity. Size--albedo correlations largely reflect 
magnitude-limited selection bias. Only 2 of 63 families survive 
diameter-limited controls. These results indicate that many apparent 
compositional trends in family-level analyses using NEOWISE albedos are 
statistically indistinguishable from measurement scatter and observational 
selection effects within current NEOWISE uncertainties.

Our detection limits and bias tests provide quantitative criteria for deciding 
when compositional inferences drawn from NEOWISE family albedos are 
statistically defensible.
\end{abstract}




\maketitle

\section{Introduction}
\label{sec:intro}

Asteroid families---groups of asteroids sharing common orbital elements 
from the collisional disruption of parent bodies \citep{Hirayama1918, 
Zappala1990, Nesvorny2015}---offer unique opportunities to study asteroid 
interior compositions. The albedo distributions of family members potentially 
encode information about parent body differentiation, compositional 
heterogeneity, and post-disruption evolutionary processes such as space 
weathering \citep{Vernazza2009, Masiero2013, DeMeo2014}.

Previous studies have generally found that asteroid families exhibit 
homogeneous albedo distributions, with deviations often attributed to 
interloper contamination---asteroids from other populations misidentified 
as family members \citep{Parker2008, Slyusarev2017, Novakovic2022, 
Chiorny2023, Erasmus2020}. Notable 
exceptions include families derived from differentiated parent bodies, 
such as Vesta, and the well-studied Nysa-Polana complex, which shows 
clear bimodality reflecting its composite origin \citep{Cellino2001, 
Masiero2013}. Some studies have reported albedo bimodality in connection 
with family ages \citep{Spoto2015} or young family identification 
\citep{Carruba2024}, and at the broader main-belt scale \citep{Usui2013}, 
while comprehensive family-level cataloging has not consistently detected 
bimodality \citep{Masiero2011}.

The \neowise{} mission has provided thermal infrared measurements for over 
100,000 asteroids \citep{Mainzer2011, Masiero2011}, enabling systematic 
studies of albedo distributions across asteroid families. However, the 
NEOWISE thermal modeling pipeline involves multiple band combinations 
depending on data availability. While 4-band observations enable full 
NEATM (Near-Earth Asteroid Thermal Model) modeling, these constitute only $\sim$2\% of results. Most asteroids 
are analyzed with 2-band or single-band data using simplified thermal 
models \citep{Mainzer2011, Myhrvold2018a}. Furthermore, several studies 
have identified systematic and random errors in NEOWISE diameters and 
albedos that may impact compositional interpretations \citep{Hanus2015, 
Myhrvold2018b, AliLagoa2017, Myhrvold2022, Moeyens2020, Whittaker2023, 
Myers2024, Myers2025}. These uncertainties are large enough that they can 
easily obscure real compositional trends, or in some cases create the 
illusion of trends that are not there.

This study addresses several questions:
\begin{enumerate}
    \item What fraction of asteroid families show bimodal albedo distributions, 
    and what is our detection efficiency?
    
    \item What systematic effects---measurement errors, selection biases, 
    population mixing---create apparent trends?
    
    \item How do observed spreads compare to measurement uncertainties, 
    and what can we reliably infer about intrinsic compositional diversity?
    
    \item How well do \neowise{} measurements agree with independent 
    \akari{} thermal modeling, and what does this imply for our error 
    budget and bimodality detection limits?
\end{enumerate}

Our goal is not to make definitive claims about parent body compositions. 
Instead, we want to establish the statistical baselines needed for family-level 
interpretations of NEOWISE albedos. Narrow albedo spreads could be misinterpreted 
as compositional homogeneity, and size-albedo correlations could be attributed 
to space weathering, when in fact measurement errors and selection biases can 
produce similar signatures. By quantifying these effects, we establish detection 
limits, bias controls, and error budgets that define what can and cannot be 
reliably inferred from NEOWISE albedos in family studies.

The remainder of this paper is organized as follows. Section~\ref{sec:data} 
describes the datasets and analysis methods. Section~\ref{sec:results} 
presents results on bimodality detection, selection bias in size--albedo 
correlations, and the relationship between observed and intrinsic albedo 
spreads. Section~\ref{sec:validation} cross-validates these findings against 
\akari{} thermal modeling and SDSS taxonomic classifications. 
Section~\ref{sec:discussion} discusses methodological limitations, a 
logarithmic-space robustness check, and recommendations for future 
\neowise{}-based family studies. Section~\ref{sec:conclusions} concludes.

\section{Data and Methods}
\label{sec:data}

\subsection{Datasets}

\subsubsection{Asteroid Families}
\label{sec:asteroid_families}

Family memberships and synthetic proper orbital elements are taken from 
the PDS Small Bodies Node\footnote{\url{https://sbn.psi.edu/pds/resource/nesvornyfam.html}}, 
which hosts both the 2015 family catalog of \citet{Nesvorny2015} and the 
2024 V2.0 update \citep{Nesvorny2024}. We combined both catalogs and 
retained families with $\geq$30 members having valid \neowise{} albedo 
measurements ($n$ denotes sample size throughout), yielding 102 families 
total (82 from 2015, 20 from 2024). The minimum threshold of 30 members 
ensures sufficient statistics for kernel density estimation (KDE)-based 
bimodality detection.

\subsubsection{NEOWISE Albedos}

V-band geometric albedos are derived from \neowise{} W1 (3.4~$\mu$m) and 
W2 (4.6~$\mu$m) observations using the NEATM thermal model \citep{Mainzer2011, 
Masiero2021}. The NEOWISE Reactivation mission uses a single thermal band 
(W2), requiring a fixed beaming parameter ($\eta = 0.95 \pm 0.2$) for the 
Main Belt \citep{Masiero2021}. Observations where W2 flux includes $>10\%$ 
reflected sunlight are filtered to ensure reliable thermal modeling 
\citep{Masiero2021}. This removes ~29\% of Main Belt observations. 
High-albedo objects are more likely to exceed the 10\% threshold due to 
stronger reflected light relative to thermal emission. Our sample includes 
124,091 asteroids with $0 < p_V < 1$. The mean reported uncertainty is 
\sigerr{} $= 0.041$.

\subsubsection{AKARI Albedos}

The \akari{} Asteroid Catalog \citep{Usui2011} provides independent thermal 
measurements at 9 and 18~$\mu$m---wavelengths that sample purely thermal 
emission without reflected light contamination. We matched 1{,}498 asteroids 
between the \akari{} and \neowise{} catalogs. The family-level cross-validation 
(Section~\ref{sec:akari}) uses the subset belonging to Nesvorn\'y families.

\subsubsection{SDSS Photometric Classifications}

Taxonomic types from \sdss{} multi-band photometry \citep{Sergeyev2021} are 
available for over 56,000 asteroids, including classification probabilities 
that we use to evaluate classification reliability. The photometric 
classification follows the scheme of \citet{DeMeo2013}, building on the 
original SDSS Moving Object Catalog \citep{Ivezic2001}. We note that these 
are color-based classifications from broadband photometry, not spectroscopic 
observations.

\subsection{Analysis Methods}

\subsubsection{Bimodality Detection}
\label{sec:bimodality_detection}

We use four criteria to define genuine bimodality, all of which must be satisfied:
\begin{enumerate}
    \item Sarle's bimodality coefficient \citep{SAS1990} $b = (\gamma^2 + 1)/(\kappa + 3) > 0.556$, 
    where $\gamma$ is skewness and $\kappa$ is excess kurtosis
    \item KDE-based valley prominence $> 0.25$
    \item Secondary peak height $> 20\%$ of primary peak
    \item Peak separation $> 1.2\sigma$, where $\sigma$ is the family 
    $\log_{10} p_V$ standard deviation
\end{enumerate}

Our primary bimodality analysis is performed on the base-10 logarithm of the 
geometric albedo, $\log_{10} p_V$, rather than on $p_V$ directly. Asteroid 
albedos are approximately log-normal \citep{Wright2016}, and the dark and bright 
compositional populations (primitive/C-complex near $p_V \approx 0.05$ and 
silicaceous/S-complex near $p_V \approx 0.25$) are more nearly symmetric and 
comparably dispersed in $\log_{10} p_V$ than in linear $p_V$, where the bright 
component carries a much longer tail. Working in log space therefore helps in
two ways. It makes the two-Gaussian mixture model used in our
detection-efficiency simulations (Section~\ref{sec:detection_results}) a better
fit, and it lets the four criteria detect the asymmetric dark/bright mixtures
that the moment-based Sarle coefficient tends to miss in linear 
space. The criteria are applied to $\log_{10} p_V$ throughout. Linear-space
results are reported in Section~\ref{sec:log10} as a robustness cross-check. A
$\Delta$BIC $> 10$ (Bayesian Information Criterion; two- vs.\ one-component 
Gaussian mixture) indicates strong evidence for two populations \citep{Kass1995}
and is required for the genuine classification of families that fail the Sarle
criterion.

Sarle's coefficient \citep{SAS1990} provides a simple moment-based test 
for bimodality, but it has a known weakness. It assumes symmetric bimodality 
and misses asymmetric two-component mixtures \citep[cf.][]{Ashman1994}. 
We tested Hartigan's dip test \citep{Hartigan1985} on several large families 
(Phocaea, Gefion, Themis, Eos, Flora; n=748--3953) but found no significant 
deviations from unimodality (all p$>$0.09), despite clear visual evidence of 
two-peak structure in KDE analysis (see Appendix~\ref{app:hartigan} for 
detailed results). 

We independently validate these classifications with four formal modality tests
from the R \texttt{multimode} package \citep{Ameijeiras2021}; the procedure and
results are reported in Section~\ref{sec:obs_rate} (Table~\ref{tab:multimode}).

For KDE, we use Scott's rule for bandwidth selection \citep{Scott1979, Scott1992}. 
Parameter sensitivity is assessed in Appendix~\ref{app:kde}. 
Valley prominence is simply $V = 1 - d_{\rm valley}/d_{\rm avg}$, where 
$d_{\rm valley}$ is the minimum density between the two highest peaks and 
$d_{\rm avg}$ is their average height. We identify peaks in the smoothed KDE 
(Gaussian filter with $\sigma = 3$ grid points) using a 5\% prominence threshold.

These numerical thresholds (Sarle $> 0.556$, valley prominence $> 0.25$, 
secondary peak $> 20\%$, separation $> 1.2\sigma$) are operational choices 
calibrated to avoid false positives under \neowise{}-like noise. The Sarle cut 
is the exception with a distributional basis, since $b = 5/9 \approx 0.556$ is 
the coefficient's value for a uniform distribution, so larger values indicate a 
distribution flatter-topped or more two-peaked than uniform 
\citep{SAS1990, Ashman1994}. The remaining, density-based thresholds are not 
physically motivated boundaries.

We use kernel density estimation (KDE) rather than Gaussian Mixture Models 
(GMM) because GMM always fits two components even to unimodal data, potentially 
creating misleading valleys. BIC analysis \citep{Schwarz1978} 
confirms this concern. It favors two-component models for skewed unimodal 
distributions (e.g., Themis, $\Delta$BIC $= +247$) despite the absence of a second 
KDE peak. Our secondary peak criterion ensures we detect genuine modes, not just 
statistical fluctuations or skewness.

Families meeting the Sarle criterion ($> 0.556$) but lacking sufficient structural 
support are classified as ``Weak Bimodal'' (Table~\ref{tab:bimodality}). 
We need one further category only for the linear-space cross-check 
(Section~\ref{sec:log10}). In linear $p_V$, some families \textit{fail} the
Sarle criterion but still show clear two-peak KDE structure. We flag these as
``Hidden Bimodal'' when they meet tighter thresholds (valley prominence $> 0.3$,
separation $> 1.5\sigma$) and have parametric support ($\Delta$BIC $> 10$,
Section~\ref{sec:bic}). In our primary $\log_{10} p_V$ classification these same
families pass or fail the Sarle criterion directly and are classified as genuine
or weak, so the Hidden Bimodal class does not appear in log space
(Table~\ref{tab:bimodality}). In the linear cross-check, the tighter thresholds
make up for the missing moment-based confirmation and reduce skewness-driven
false positives.

\label{sec:bic}
This same $\Delta$BIC $> 10$ threshold also serves as a required consistency 
check for families that fail the Sarle criterion but exhibit KDE two-peak 
structure. Before assigning genuine status in log space or ``Hidden Bimodal'' 
status in the linear cross-check (Section~\ref{sec:log10}), we require 
parametric support from a Bayesian Information Criterion comparison between 
one- and two-component Gaussian mixtures. This guards against a 
KDE-identified two-peak structure being strongly rejected by the parametric 
model comparison.

\subsubsection{Size-Albedo Correlations}

We test for size-albedo correlations using Spearman rank correlation 
($\rho$), which is robust to outliers and non-linear relationships. 
Statistical significance is assessed via two-tailed tests, with $p$-values 
indicating the probability of obtaining the observed correlation under 
the null hypothesis of no correlation. We report correlations for both 
the full sample (all diameters) and diameter-limited subsamples 
($D > 5$~km) to identify selection bias effects.

\subsubsection{Detection Efficiency and False-Positive Simulations}
\label{sec:fp_sims}

To quantify detection efficiency across parameter space, we perform a systematic 
grid scan in $\log_{10} p_V$ over peak separation ($\Delta\mu = 0.10$--$0.65$ 
dex) and per-component intrinsic scatter ($\sigma_{\rm int} = 0.05$--$0.25$ dex). 
For each configuration we generate 500 independent synthetic families, each a 
500-object realisation of a two-component log-normal mixture with the dark 
component at $\mu_1 = \log_{10}(0.06) \approx -1.22$ and the bright component at 
$\mu_2 = \mu_1 + \Delta\mu$. We test both 
symmetric ($w_1{:}w_2 = 50{:}50$) and asymmetric ($w_1{:}w_2 = 15{:}85$) 
population weights. \neowise{}-like measurement errors (linear $\sigma \approx 
0.04$) are applied in linear $p_V$ and then transformed to log space, faithfully 
reproducing the albedo-dependent log-space error. We then apply our 
four-criterion detection threshold. Results are presented in 
Section~\ref{sec:detection_results}.

A complementary concern is whether measurement errors could transform genuinely 
unimodal distributions into apparent (spurious) bimodality. To test this, we 
performed Monte Carlo simulations with 500 synthetic families per configuration, 
testing 9 unimodal distribution types (Gaussian, skewed, log-normal with 
various parameters) across \neowise{}-like error levels ($\sigma = 0.03$--$0.06$) 
and family sizes ($n = 100$--$1000$). The false positive rate for spurious 
bimodal classification was 0.1\%, with a maximum of 2.2\% in worst-case 
configurations (narrow C-type distributions with high errors). Specific 
tests matching the parameters of our asymmetric dark/bright families 
(Gefion-like, Phocaea-like, etc.) yielded false positive rates of 0.0--0.1\%. 
This confirms that our bimodal detections are unlikely to be error-induced 
artifacts.

\subsubsection{Selection Bias Test}

The $D > 5$~km threshold for testing size-albedo correlations was chosen 
based on completeness analysis. The ratio of low-albedo ($p_V < 0.1$) to 
high-albedo ($p_V > 0.2$) objects increases from $\sim$0.4 at $D < 2$~km 
to $\sim$6 at $D > 7$~km, stabilizing above 5~km. This indicates that 
magnitude-limited detection tends to miss faint (low-albedo, small-diameter) 
objects below this threshold. Additionally, 
\citet{Pravec2012} demonstrated that absolute magnitude ($H$) values 
for small asteroids are systematically overestimated (too bright), 
leading to artificially high albedo estimates. This systematic bias 
provides a physical motivation for the size cutoff beyond statistical 
completeness alone.

To test robustness, we tested alternative thresholds of 3, 5, 7, and 
10~km. Results are qualitatively stable. Correlations systematically 
weaken or reverse above all thresholds. While some families shift between 
adjacent classification categories (e.g., robust $\leftrightarrow$ candidate) 
when the threshold changes by $\pm 1$~km, no family transitions directly 
from ``robust weathering signal'' to ``no signal'' or vice versa. We prefer 
$D > 5$~km as a balance between completeness and sample size.

\subsubsection{Intrinsic Variance Estimation}

Where possible, we estimate intrinsic variance as:
\begin{equation}
    \sigma^2_{\rm int} = \sigma^2_{\rm obs} - \sigma^2_{\rm err}
\end{equation}
where $\sigma_{\rm obs}$ is the observed albedo dispersion within each family 
and $\sigma_{\rm err}$ is the mean of the individual measurement 
uncertainties reported by NEOWISE. This quantity $\sigma^2_{\rm int}$ represents 
the compositional variance remaining after accounting for measurement noise, 
and we use it to assess whether observed family spreads reflect genuine 
compositional diversity or are dominated by measurement error 
(Section~\ref{sec:albedo_spreads}; see also Section~\ref{sec:methodological_limitations} 
for discussion of the limitations of this simple variance-subtraction approach).

\neowise{} reports measurement uncertainties in linear $p_V$. We 
propagate them into log space for each object as $\sigma_{\log,i} \approx
0.434\,\sigma_{p_V,i}/p_{V,i}$, the first-order transformation of a linear error
to $\log_{10}$. We take $\sigma_{\rm err}$ to be the root-mean-square of
$\sigma_{\log,i}$ over the family, and $\sigma_{\rm obs}$ to be the standard deviation 
of $\log_{10} p_V$ within the family. We use the RMS rather than the mean error 
because $\sigma_{\rm obs}$ is itself a root-mean-square (standard-deviation) 
quantity, so the comparison and the subtraction $\sigma^2_{\rm int} = 
\sigma^2_{\rm obs} - \sigma^2_{\rm err}$ are dimensionally consistent.

This simple variance subtraction assumes 
equal measurement errors for all objects and provides a first-order estimate. For many families 
$\sigma_{\rm obs} \lesssim \sigma_{\rm err}$, making this estimate unreliable 
with \neowise{} data alone. More sophisticated approaches (hierarchical Bayesian 
modeling, size-dependent error treatment) are beyond our scope but would be 
valuable for families with larger sample sizes. Independent constraints 
(e.g., from higher-precision thermal measurements or 
spectroscopic surveys) would be needed to reliably characterize 
intrinsic compositional spreads.

\section{Results}
\label{sec:results}

\subsection{Bimodality}

\subsubsection{Observed Bimodality Rate}
\label{sec:obs_rate}

Working in $\log_{10} p_V$, 6 of 102 families with $n \geq 30$ (5.9\%) satisfy 
all four bimodality criteria: Nysa-Polana, Phocaea, Tirela, Juno, Henan, and 
Telramund (Table~\ref{tab:genuine}). All six are dark/bright mixtures, with 
component peaks near $p_V \approx 0.05$--$0.06$ (primitive) and $p_V \approx 
0.21$--$0.29$ (silicaceous). In log space their Sarle coefficients cluster 
tightly at $b \approx 0.69$--$0.71$, whereas in linear $p_V$ the same families 
span $b \approx 0.32$--$0.70$. The asymmetric dark/bright mixtures that the 
moment-based coefficient underweights in linear space become unambiguous once 
the long bright tail is removed by the log transform. Two of the six, 
Nysa-Polana and Juno, are bimodal in both representations. The other four are 
resolved specifically in log space (Section~\ref{sec:log10}). A further 15 
families have high Sarle but fail at least one structural criterion (weak bimodal; Section~\ref{sec:weak}), and the 
remainder are unimodal (Table~\ref{tab:bimodality}).

\begin{table}[!ht]
\caption{Bimodality Classification}
\label{tab:bimodality}
\centering
\begin{threeparttable}
\begin{tabular*}{\tblwidth}{@{}LCC@{}}
\toprule
Classification & $N$ & \% \\
\midrule
Genuinely Bimodal & 6 & 5.9 \\
Weak Bimodal & 15 & 14.7 \\
Wide Unimodal & 7 & 6.9 \\
Skewed Unimodal & 17 & 16.7 \\
Unimodal & 57 & 55.9 \\
\bottomrule
\end{tabular*}
\begin{tablenotes}\small
\item Classification in $\log_{10} p_V$. Genuinely bimodal requires all four 
criteria. ``Weak Bimodal'' has Sarle $> 0.556$ but fails at least one structural
	criterion.  Most lack a resolved second peak, and a few show a KDE
	valley whose secondary peak lies below the 20\% height threshold. The
	hidden-bimodal class required in linear space (clear KDE structure but
	low Sarle) does not arise in log space, where such families are
	classified directly as genuine; see Section~\ref{sec:log10}.
\end{tablenotes}
\end{threeparttable}
\end{table}

\begin{figure*}
\centering
\includegraphics[width=\textwidth]{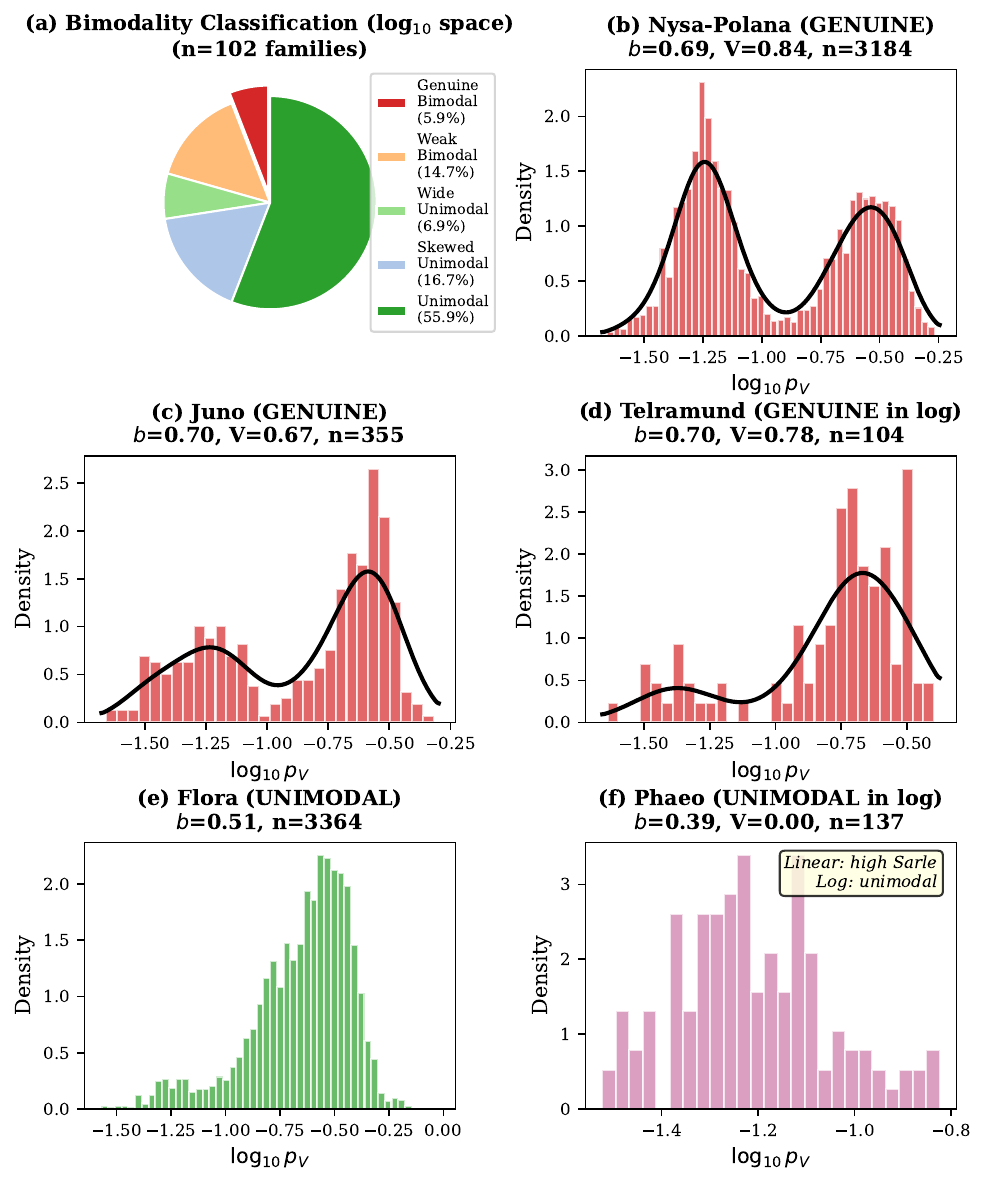}
\caption{Bimodality analysis of 102 asteroid families in $\log_{10} p_V$. 
(a) Classification summary: 6 genuinely bimodal (5.9\%), 15 weak bimodal 
(high Sarle but incomplete structural support, 14.7\%), and 81 unimodal-class families (wide/skewed/ 
unimodal). (b) Nysa-Polana and (c) Juno, genuinely bimodal in both linear and 
log space. (d) Telramund, resolved as genuinely bimodal in log space 
($b \approx 0.70$) though classified wide-unimodal in linear $p_V$. (e) Flora, a 
representative unimodal S-complex family. (f) Phaeo, whose high linear Sarle 
reflects skewness alone and resolves to unimodal in log space. All 
distributions are shown in $\log_{10} p_V$.}
\label{fig:bimodality}
\end{figure*}

The six genuinely bimodal families are listed in Table~\ref{tab:genuine}. 
Figure~\ref{fig:bimodality} illustrates the classification and representative 
$\log_{10} p_V$ distributions for families genuinely bimodal in both 
representations (Nysa-Polana, Juno) and one that is genuine only in log space 
(Telramund).

\begin{table*}[!ht]
\caption{Genuinely Bimodal Families}
\label{tab:genuine}
\centering
\begin{threeparttable}
\begin{tabular*}{\tblwidth}{@{}LCCCCCCCCC@{}}
\toprule
Family & $n$ & Sarle & Valley & 2nd/1st & Sep ($\sigma$) & $\mu_1$ & $\mu_2$ & $\Delta\mu$ & $w_1{:}w_2$ \\
\midrule
Nysa-Polana & 3184 & 0.69 & 0.84 & 0.74 & 1.95 & $-1.24$ & $-0.56$ & 0.68 & 56:44 \\
Phocaea & 925 & 0.71 & 0.89 & 0.27 & 2.20 & $-1.24$ & $-0.54$ & 0.70 & 21:79 \\
Tirela & 376 & 0.70 & 0.72 & 0.31 & 2.26 & $-1.24$ & $-0.64$ & 0.60 & 26:74 \\
Juno & 355 & 0.70 & 0.67 & 0.50 & 1.86 & $-1.26$ & $-0.61$ & 0.65 & 40:60 \\
Henan & 187 & 0.71 & 0.69 & 0.28 & 2.33 & $-1.22$ & $-0.65$ & 0.57 & 25:75 \\
Telramund & 104 & 0.70 & 0.78 & 0.23 & 2.39 & $-1.37$ & $-0.67$ & 0.70 & 17:83 \\
\bottomrule
\end{tabular*}
\begin{tablenotes}\small
\item Genuinely bimodal families in $\log_{10} p_V$. Columns: $n$ = members; 
Sarle = bimodality coefficient; Valley = KDE valley prominence; 2nd/1st = ratio 
of secondary to primary KDE peak height; Sep = peak separation in units of the 
family $\log_{10} p_V$ standard deviation; $\mu_1,\mu_2$ = two-component 
Gaussian-mixture means in $\log_{10} p_V$ (dark, bright); 
$\Delta\mu = |\mu_2-\mu_1|$; $w_1{:}w_2$ = component weights. Weights range from 
near-balanced (Nysa-Polana, Juno) to strongly bright-dominated (Phocaea, 
Telramund). Nysa-Polana is a complex of overlapping families (Nysa, Polana, 
Eulalia, Hertha; \citealt{Walsh2013,Dykhuis2015}), so its bimodality reflects 
mixing of distinct parent bodies; population-mixing and interloper explanations 
apply to all six and not only to the linear-space detections 
(cf.\ Section~\ref{sec:juno}).
\end{tablenotes}
\end{threeparttable}
\end{table*}

We validated our classification using Ameijeiras-Alonso, Crujeiras \& 
Rodríguez-Casal's R \texttt{multimode} package \citep{Ameijeiras2021}, accessed from 
Python via the \texttt{rpy2} bridge. For each family we ran four tests at 
both $\mathrm{mod}_0 = 1$ and $\mathrm{mod}_0 = 2$: Silverman's critical 
bandwidth test \citep{Silverman1981}, Hall \& York's calibrated test 
\citep{Hall2001}, the Fisher--Marron excess mass test 
\citep{Muller1991}, and the ACR test (named for the package authors' 
initials) 
\citep{Ameijeiras2019}. Hall \& York's calibration is defined only for 
testing unimodality ($\mathrm{mod}_0=1$). For tractability, families 
with $n > 500$ members were randomly subsampled to $n_{\rm 
used} = 500$ (random seed 42 for reproducibility). We use $B = 200$ 
bootstrap replicates throughout. Results 
(Table~\ref{tab:multimode}) confirm our four best-populated genuine 
families---Nysa-Polana, Phocaea, Tirela, and Juno---as bimodal. All four 
tests reject unimodality but not bimodality ($p < 0.05$ at $k=1$, 
$p > 0.05$ at $k=2$). The two smaller genuine families, Henan ($n=187$) and 
Telramund ($n=104$), are supported by the Hall--York and Fisher--Marron 
tests but not by Silverman's test or ACR, consistent with their smaller 
samples and strongly asymmetric mixtures. Control families Koronis and 
Themis are correctly identified as unimodal by all four tests ($p > 0.24$ 
for all $k=1$ tests).

\begin{table*}[!ht]
\caption{Modality Test Results}
\label{tab:multimode}
\centering
\begin{threeparttable}
\begin{tabular*}{\tblwidth}{@{}LCCCCCC@{}}
\toprule
Family & $n$ ($n_{\rm used}$) & Silverman & Hall-York & Fisher-Marron & ACR & Our Class \\
\midrule
Nysa-Polana & 3184 (500) & B & B & B & B & GENUINE \\
Phocaea     & 925  (500) & B & B & B & B & GENUINE \\
Tirela      & 376        & B & B & B & B & GENUINE \\
Juno        & 355        & B & B & B & B & GENUINE \\
Henan       & 187        & U & B & B & U & GENUINE \\
Telramund   & 104        & U & B & B & U & GENUINE \\
Gefion      & 748  (500) & B & B & B & B & WEAK \\
Rafita      & 283        & U & B & B & U & WEAK \\
Witt        & 89         & U & U & B & U & WEAK \\
Koronis     & 1014 (500) & U & U & U & U & UNIMODAL \\
Themis      & 2202 (500) & U & U & U & U & UNIMODAL \\
\bottomrule
\end{tabular*}
\begin{tablenotes}\small
\item U = Unimodal (fail to reject $k=1$ at $\alpha=0.05$); B = Bimodal 
(reject $k=1$, fail to reject $k=2$); M = Multimodal (reject both $k=1$ 
and $k=2$). HY columns at $k=2$ are not reported as Hall \& York's 
calibration is defined only for $k=1$.
\end{tablenotes}
\end{threeparttable}
\end{table*}

\subsubsection{Asymmetric bimodals resolved in log space}

\begin{figure*}
\centering
\includegraphics[width=\textwidth]{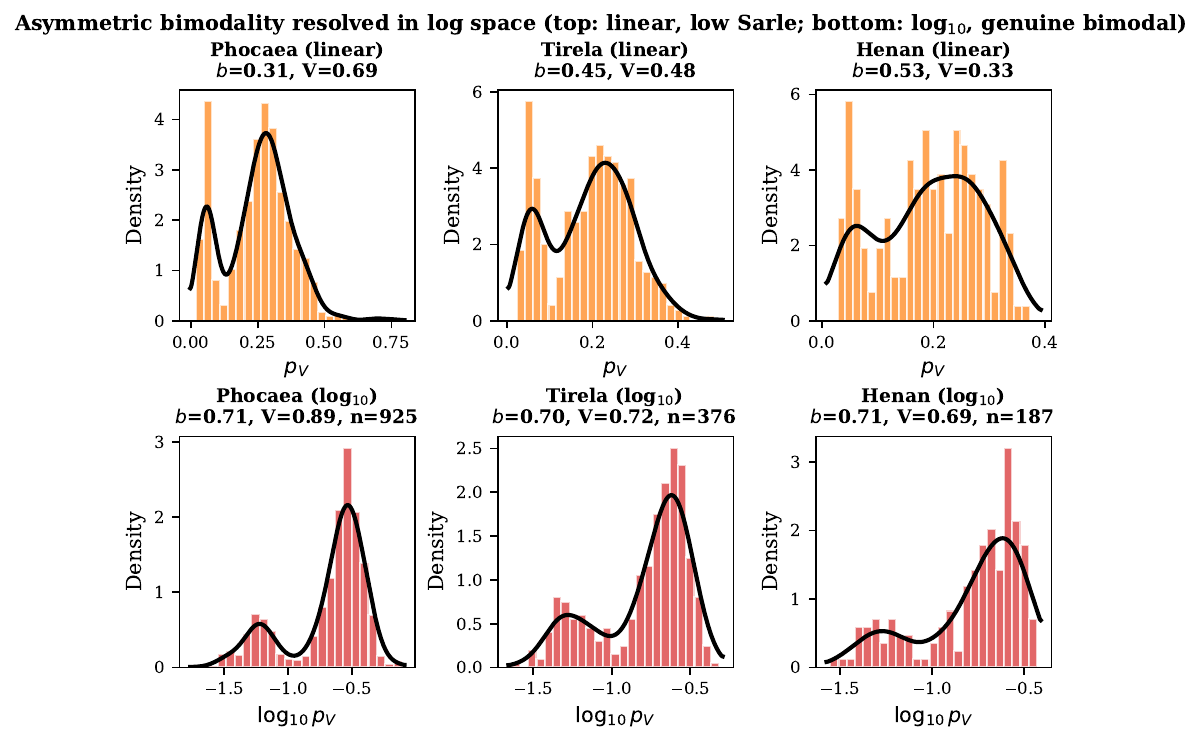}
\caption{Three asymmetric dark/bright families (Phocaea, Tirela, and 
Henan) in linear $p_V$ (top row) and $\log_{10} p_V$ (bottom row). Each panel 
shows the histogram and KDE fit with the Sarle coefficient ($b$); exact values 
are given in Table~\ref{tab:hidden}. In linear space the dark component 
appears mainly as a tail and Sarle stays below threshold. In log space the 
two components separate cleanly and all three satisfy the genuine-bimodality 
criteria.}
\label{fig:hidden}
\end{figure*}

Four of the six genuine families (Phocaea, Tirela, Henan, and Telramund) are
not flagged by Sarle's coefficient in linear $p_V$ ($b = 0.31$--$0.53$;
Table~\ref{tab:log10_comparison}); full linear-space diagnostics for three of
these (Phocaea, Tirela, Henan) are given in Table~\ref{tab:hidden}. Their dark/bright mixtures are strongly 
asymmetric, with a numerically dominant bright population
($p_V \approx 0.22$--$0.29$) and a minor dark component 
($p_V \approx 0.05$--$0.07$). The long bright tail therefore suppresses the
moment-based signal. In $\log_{10} p_V$ the same families satisfy all four 
criteria ($b \approx 0.70$; Table~\ref{tab:genuine}), with KDE valley 
prominences of 0.69--0.89 and peak separations of 2.2--2.4$\sigma$ 
(classifications robust to KDE parameter variations; Appendix~\ref{app:kde}). 
Figure~\ref{fig:hidden} shows three of them (Phocaea, Tirela, Henan) in linear 
$p_V$ (top row) and $\log_{10} p_V$ (bottom row). The second peak, marginal or 
absent to the moment test in linear space, is unambiguous in log. This is the 
main methodological reason for using the logarithmic representation 
(Section~\ref{sec:log10}).

The same statistical structure has more than one possible physical cause. It
could reflect structured interloper contamination, error-convolved tails of skewed
distributions, or incomplete dynamical mixing of distinct source regions. These 
alternatives apply to all six genuine detections, not only to these four 
asymmetric cases. The three remaining linear-hidden families (Gefion, Rafita, 
Witt) retain high Sarle in log space but lack a sufficiently resolved second 
peak and are classified weak bimodal (Table~\ref{tab:log10_comparison}).

\begin{table*}[!ht]
\caption{Asymmetric Dark/Bright Families: Linear-Space Metrics}
\label{tab:hidden}
\centering
\begin{threeparttable}
\begin{tabular*}{\tblwidth}{@{}LCCCCCCCC@{}}
\toprule
Family & $n$ & Sarle & Valley & 2nd/1st & Sep ($\sigma$) & $\mu_1$ & $w_1:w_2$ & $\Delta$BIC \\
\midrule
Gefion & 748 & 0.34 & 0.76 & 0.37 & 2.10 & 0.054 & 13:87 & +198 \\
Phocaea & 925 & 0.31 & 0.69 & 0.61 & 1.75 & 0.057 & 19:81 & +340 \\
Witt & 89 & 0.46 & 0.60 & 0.32 & 2.07 & 0.062 & 15:85 & +12 \\
Rafita & 283 & 0.38 & 0.60 & 0.35 & 1.86 & 0.066 & 14:86 & +34 \\
Tirela & 376 & 0.45 & 0.48 & 0.71 & 1.78 & 0.056 & 22:78 & +116 \\
Henan & 187 & 0.53 & 0.33 & 0.66 & 1.92 & 0.053 & 18:82 & +45 \\
\bottomrule
\end{tabular*}
\begin{tablenotes}\small
\item Linear-space metrics for the six asymmetric dark/bright families; their 
low linear Sarle reflects the asymmetric mixture. Columns: $n$ = members; 
Sarle = bimodality coefficient; Valley = KDE valley prominence; 2nd/1st = 
ratio of secondary to primary KDE peak height; Sep = peak separation in 
units of the family standard deviation; $\mu_1$ = dark-component mean; 
$w_1{:}w_2$ = component weights. Log-space classification of each family 
(genuine vs.\ weak bimodal) is given in 
Table~\ref{tab:log10_comparison}. $\Delta$BIC $> 10$ indicates strong 
preference for a two-component model; Massalia was excluded ($\Delta$BIC $= -12$).
\end{tablenotes}
\end{threeparttable}
\end{table*}

In every case the bright population dominates numerically (weights 74--83\%; 
Table~\ref{tab:genuine}), so these are dominant populations with a 
$\sim$17--26\% dark component rather than balanced mixtures like Nysa-Polana. 
The dark component may represent primordial heterogeneity in the parent body, 
C-type interlopers from dynamical mixing, or family-membership 
misclassification. Distinguishing these requires independent spectral data.

\subsubsection{Detection Efficiency}
\label{sec:detection_results}

\begin{figure*}
\centering
\includegraphics[width=\textwidth]{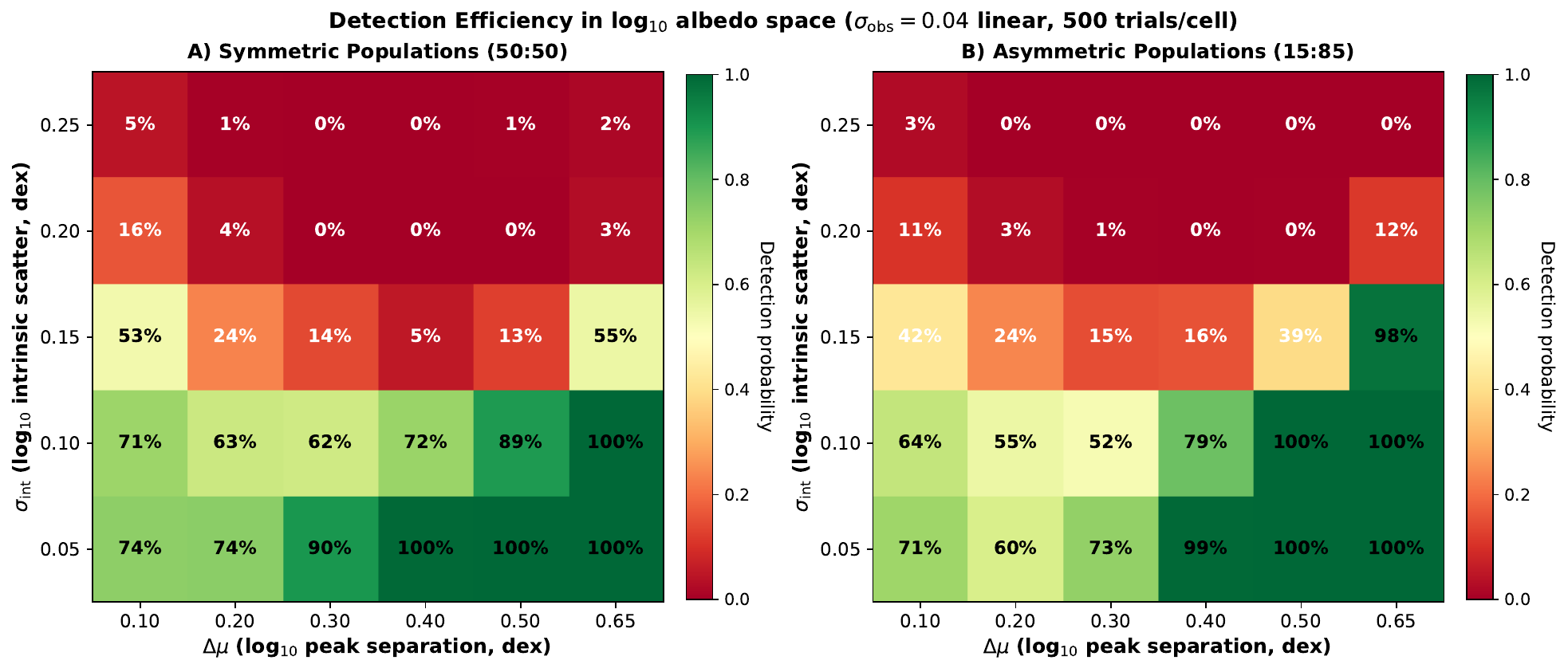}
\caption{Bimodality detection efficiency in $\log_{10} p_V$ from Monte Carlo 
simulations of synthetic families. The grid varies log peak separation 
($\Delta\mu = 0.10$--$0.65$ dex) and per-component intrinsic scatter 
($\sigma_{\rm int} = 0.05$--$0.25$ dex). (a) Symmetric populations 
($w_1{:}w_2 = 50{:}50$). (b) Asymmetric populations ($w_1{:}w_2 = 15{:}85$). 
Numerical values within cells give the detection percentage from 500 
simulations. The characteristic dark/bright separation is $\Delta\mu \approx 
0.64$ dex (rightmost column).}
\label{fig:detection_grid}
\end{figure*}

To estimate how many bimodal families our method might miss, we perform 
systematic parameter space simulations (Figure~\ref{fig:detection_grid}). 
Results reveal strong dependence on both peak separation and population asymmetry. 
At the separation characteristic of dark/bright (C/S) mixtures 
($\Delta\mu \approx 0.64$ dex), detection rises to $\sim$100\% when the 
per-component intrinsic scatter is small ($\sigma_{\rm int} \lesssim 0.10$ dex) 
and falls to $<$12\% once $\sigma_{\rm int} \gtrsim 0.20$ dex, for both 
symmetric and asymmetric (15:85) configurations. The false-positive rate 
established by the unimodal-control simulations (Section~\ref{sec:fp_sims}; 
0.1\%, rising to 2.2\% only in worst-case high-scatter configurations) confirms 
conservative classification. Bootstrap stability analysis 
(Appendix~\ref{app:bootstrap}) reveals median classification stability of 
77\%, with the six genuine families showing 91\% mean stability 
(individually 67--100\%).

This detectability ceiling ($\sigma_{\rm int} \approx 0.15$--$0.18$ dex at the 
characteristic separation) sits above the intrinsic scatter present in the 
families. The family-averaged observed log dispersion is $\sigma_{\rm obs} 
\approx 0.16$ (C-complex) to $0.21$ (S-complex), and because measurement error 
accounts for nearly all of it (median $\sigma_{\rm obs}/\sigma_{\rm err} 
\approx 0.99$; Section~\ref{sec:spreads}), the intrinsic per-component scatter 
is small. The genuine families thus lie inside the regime where a dark/bright 
bimodality at the characteristic separation would be detected with high 
probability, so the scarcity of genuine bimodality (six of 102 families) 
reflects genuine compositional structure, not limited sensitivity. The elevated 
detection at very small separations combined with large $\sigma_{\rm int}$ 
(lower-left of Figure~\ref{fig:detection_grid}) does not represent genuine 
resolving power. There, the broad scatter occasionally produces noise-induced 
apparent structure, so the high-$\sigma_{\rm int}$ rows are treated as an 
extreme edge case. Families with intrinsically lower albedo contrast (e.g., 
X-complex, K-type) may still remain undetectable within the \neowise{} error 
regime.

The genuine bimodality rate in log space is therefore 5.9\% (6 of 102 
families), with no separate ``hidden'' category required.

Our 15:85 mixture ratio represents the most asymmetric configuration we can 
reliably probe at \neowise{} noise levels. Even at this ratio, detection 
exceeds $\sim$98\% at the characteristic separation ($\Delta\mu \approx 0.65$ 
dex) for $\sigma_{\rm int} \leq 0.10$ dex, but collapses below $\sim$15\% once 
$\sigma_{\rm int} \geq 0.20$ dex (Figure~\ref{fig:detection_grid}b). More 
extreme asymmetries (e.g., $w_1{:}w_2 \approx 5{:}95$ or lower) would require 
deeper or higher-precision surveys to detect. We discuss this implication in 
Section~\ref{sec:recommendations}.

\subsubsection{Weak-bimodal families}
\label{sec:weak}

Fifteen families are classified weak bimodal in $\log_{10} p_V$
(Table~\ref{tab:bimodality}). They satisfy the Sarle criterion ($b = 0.57$--$0.74$)
but fail at least one of the structural criteria (Section~\ref{sec:bimodality_detection}),
and we do not treat them as two-population systems. Only three---Gefion,
Massalia, and Chimaera---show a KDE valley in log space at all, and in all three
the secondary peak lies below the 20\% height threshold (secondary-to-primary
ratios of $0.12$--$0.18$), so none has a fully resolved second component and none is promoted to genuine. The remaining twelve
show no resolved second peak at all. Their elevated Sarle coefficients arise from
distribution skewness and heavy tails ($|\gamma| = 0.4$--$2.9$ in log) rather than
from a distinct dark or bright component. The independent modality tests point the
same way. Among the weak families only Gefion is flagged bimodal by all four
\texttt{multimode} tests---consistent with its KDE valley and sub-threshold secondary peak---whereas Rafita and Witt are predominantly unimodal
(Table~\ref{tab:multimode}). These fifteen families originate from a range of
linear-space classes (seven unimodal, three ``hidden'', three weak, and two skewed
unimodal), so the weak-bimodal category is not an artifact of either representation
but the buffer that absorbs skewed single populations and marginal cases. Its
existence is what keeps the genuine class conservative. The full list, with
log-space diagnostics and the linear-space classification of each family, is given
in Appendix~\ref{app:weak} (Table~\ref{tab:weak}).

\subsection{Population Mixing and Selection Bias: Juno as Case Study}
\label{sec:juno}

\begin{figure*}
\centering
\includegraphics[width=\textwidth]{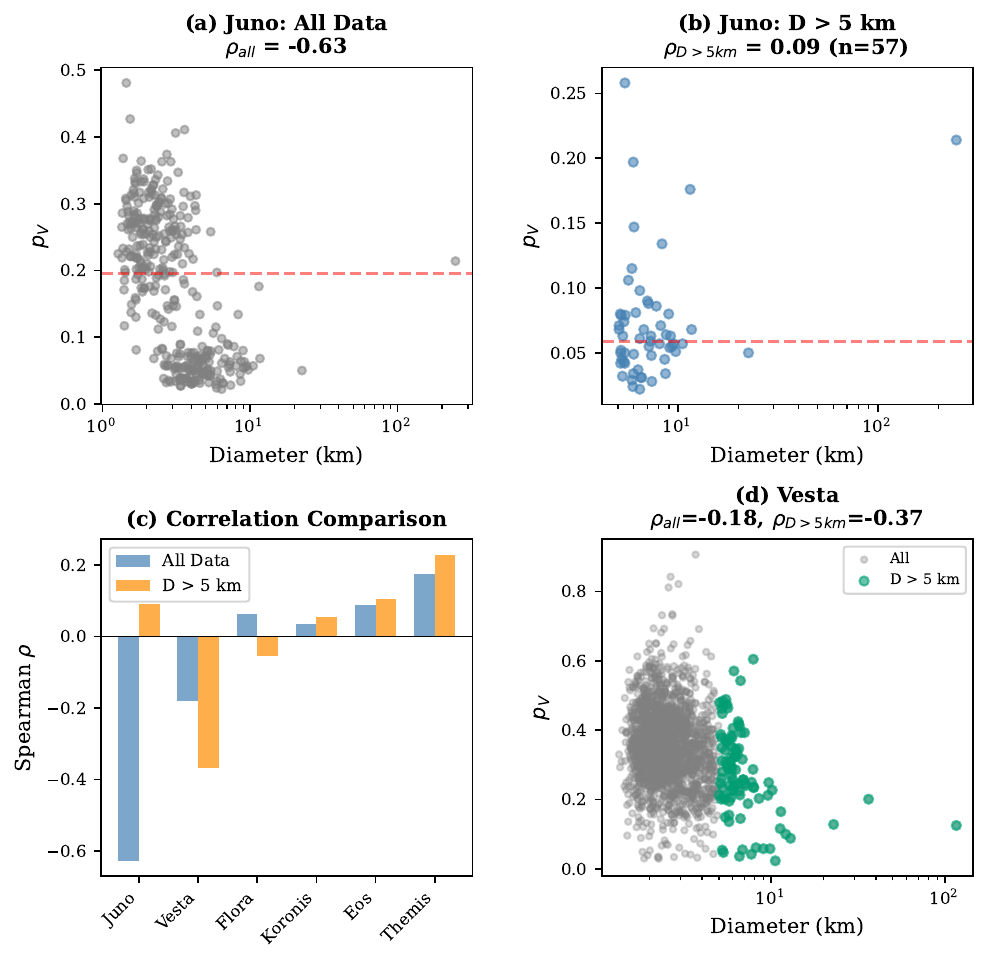}
\caption{Selection bias in size--albedo correlations, illustrated using the 
Juno family. (a) Global Spearman correlation, $\rho = -0.63$. 
(b) Restricted to $D > 5$~km, $\rho = +0.09$. (c) Comparison across multiple 
families: global correlations (blue) vs $D > 5$~km (orange). (d) Vesta 
size--albedo relation: unlike Juno, the correlation \textit{strengthens} when 
restricted to $D > 5$~km ($\rho_{\rm global} = -0.18 \to \rho_{D>5\,\rm km} = 
-0.37$), opposite to the selection-bias trend, marking it as a weathering 
candidate.}
\label{fig:selection}
\end{figure*}

We present Juno as an illustrative case study of how population mixing 
and selection bias can create spurious correlations (Figure~\ref{fig:selection}). 
Juno shows a strong global size-albedo correlation ($\rho = -0.63$) that 
might be interpreted as space weathering evidence.

However, this correlation has two non-physical origins:

\textbf{Population mixing:} Juno contains two nearly equal albedo populations 
with different size distributions. When analyzed separately, neither 
population shows significant correlation (Table~\ref{tab:juno}).

\textbf{Selection bias:} Magnitude-limited detection favors high-albedo 
objects at small sizes. Restricting to $D > 5$~km---where detection is 
more complete---reverses the correlation to $\rho = +0.09$.

\begin{table}[!ht]
\caption{Juno Family: Correlation Analysis}
\label{tab:juno}
\centering
\begin{threeparttable}
\begin{tabular*}{\tblwidth}{@{}LCCC@{}}
\toprule
Sample & $n$ & $\rho$ & Interpretation \\
\midrule
Global & 355 & $-0.63$ & Apparent correlation \\
Population 0 & 178 & $-0.39$ & Reduced \\
Population 1 & 177 & $-0.02$ & None \\
$D > 5$~km & 57 & $+0.09$ & Selection bias removed \\
\bottomrule
\end{tabular*}
\begin{tablenotes}\small
\item $\rho$ = Spearman rank correlation between diameter and $p_V$; $n$ = 
sample size. Population~0 and Population~1 are the two albedo subpopulations 
identified within Juno (Section~\ref{sec:juno}); $D > 5$~km is the full 
sample restricted to the diameter-limited cut.
\end{tablenotes}
\end{threeparttable}
\end{table}

Note that Juno may represent an extreme case. Other bimodal families 
(e.g., Nysa-Polana) retain correlations even after restricting to $D > 5$~km 
($\rho = -0.31$), suggesting some may reflect genuine physical processes. 
Nevertheless, this example demonstrates the importance of testing for 
these effects before interpreting correlations as evidence for space weathering.

\subsubsection{Weathering Signatures in Unimodal Families}

We extend the selection bias test to all families with sufficient data. 
For each family with $>50$ members and $\geq 20$ objects with $D > 5$~km 
(63 families total), we compare the global size-albedo Spearman correlation 
coefficient ($\rho_{\rm global}$) to the correlation restricted to large 
objects ($\rho_{D>5\rm km}$). A ``robust'' weathering signature requires 
all four criteria: (1) $\rho_{\rm global} < -0.2$, (2) $\rho_{D>5\rm km} < -0.15$, 
(3) $n_{D>5\rm km} \geq 20$, and (4) $p_{\rm global} < 0.01$ (two-tailed 
$p$-value for the global correlation).

The requirement that $|\rho_{\rm global}| > |\rho_{D>5\rm km}|$ is not 
explicitly enforced but is expected if selection bias drives the correlation: 
small asteroids amplify apparent trends due to magnitude-limited sampling. 
If the correlations were equal, the trend would be uniform across sizes, 
suggesting a real physical effect. Vesta shows the opposite pattern 
($\rho_{D>5\rm km} = -0.37$ vs $\rho_{\rm global} = -0.18$), strengthening 
at large diameters, which merits further investigation.

Of 63 tested families, only 2 pass all four criteria: Nysa-Polana 
($\rho_{\rm global} = -0.40$, $\rho_{D>5\rm km} = -0.31$) and Phaeo 
($\rho_{\rm global} = -0.32$, $\rho_{D>5\rm km} = -0.20$). Nysa-Polana 
is a well-known bimodal family where the correlation likely reflects 
population mixing. Phaeo (n=137, X-type) has a high \emph{linear} Sarle coefficient 
(0.57) arising from skewness, but shows no KDE valley structure (valley 
prominence = 0) and resolves to unimodal in log space 
(Figure~\ref{fig:bimodality}f). As a genuinely unimodal family, it is the only 
unimodal family passing our robust weathering test, marking it as a weathering 
candidate worth follow-up.

Several families show statistically significant global correlations 
(e.g., Vesta: $\rho_{\rm global} = -0.18$, $p < 10^{-15}$), but fail 
Criterion 1 ($\rho_{\rm global} > -0.2$). Vesta's 
correlation \textit{strengthens} at large diameters ($\rho_{D>5\rm km} = -0.37$), 
opposite to selection bias expectations, making it a compelling 
weathering candidate despite failing our conservative threshold.

Note that strengthening correlations at large diameters 
can also arise from ``population filtering''---if a family contains an 
undetected low-albedo subpopulation that is more likely to be lost at 
large sizes, the remaining high-albedo members may show artificially 
strong correlations. We verified that Vesta has minimal low-albedo 
contamination ($<0.5\%$ outliers, all high-albedo), supporting a 
genuine weathering interpretation. Other families with similar 
patterns (e.g., Watsonia) have wider albedo distributions that suggest
unresolved population structure and need careful interpretation. Telramund 
shows a comparable pattern and is in fact genuinely bimodal in log space 
(Section~\ref{sec:results}, Table~\ref{tab:genuine}).

In most families, correlations weaken or reverse at large diameters, 
suggesting that apparent size-albedo correlations in \neowise{} data 
are predominantly artifacts of magnitude-limited selection.

\subsubsection{Exploratory Analysis: Candidate Weathering and Differentiation Signals}

While no unimodal family unambiguously passes our conservative four-criterion 
test, several show suggestive signals that merit further study 
with independent data. Vesta, with its strengthening correlation at large 
diameters, is a notable exception candidate. We present these as 
\textit{exploratory findings} with explicit caveats about selection bias 
contamination (Tables~\ref{tab:weathering_candidates} and \ref{tab:differentiation}).

\begin{table*}[!ht]
\caption{Candidate Weathering Signals in Unimodal Families}
\label{tab:weathering_candidates}
\centering
\begin{threeparttable}
\begin{tabular*}{\tblwidth}{@{}LRCRRRL@{}}
\toprule
Family & $n$ & Type & $\rho_{\rm global}$ & $\rho_{D>5km}$ & $p$ & Caveat \\
\midrule
Vesta & 1891 & V & $-0.18$ & $-0.37$ & $< 10^{-15}$ & $\rho_{\rm global} > -0.2$ \\
Gersuind & 201 & D & $-0.21$ & --- & $3 \times 10^{-3}$ & $n_{D>5km} < 20$ \\
Maria & 892 & S & $-0.11$ & $-0.11$ & $1 \times 10^{-3}$ & Weak $\rho$ \\
Euphrosyne & 1520 & C & $-0.12$ & --- & $2 \times 10^{-6}$ & Weak $\rho$ \\
\bottomrule
\end{tabular*}
\begin{tablenotes}\small
\item Families with significant global correlations ($p < 0.01$) but failing 
the robust four-criterion test. ``---'' indicates insufficient sample at 
$D > 5$~km.
\end{tablenotes}
\end{threeparttable}
\end{table*}

We also tested for size-dependent composition by comparing mean albedos 
of small ($D < 5$~km) versus large ($D > 5$~km) members 
(Table~\ref{tab:differentiation}). Significant differences could indicate 
parent body differentiation \citep{Weiss2012}, though selection bias remains a concern.

\begin{table*}[!ht]
\caption{Size-Dependent Albedo: Candidate Differentiation Signals}
\label{tab:differentiation}
\centering
\begin{threeparttable}
\begin{tabular*}{\tblwidth}{@{}LRCRRRL@{}}
\toprule
Family & $n$ & Type & $\langle p_V \rangle_{\rm small}$ & $\langle p_V \rangle_{\rm large}$ & $\Delta p_V$ & Note \\
\midrule
\multicolumn{7}{c}{\textit{Unimodal Families}} \\
\midrule
Vesta & 1891 & V & 0.365 & 0.310 & $+0.055$ & Weathering? \\
Eunomia & 2166 & S & 0.267 & 0.247 & $+0.020$ & Weathering? \\
Flora & 3364 & S & 0.240 & 0.255 & $-0.015$ & Unexpected \\
Eos & 3953 & S & 0.146 & 0.159 & $-0.013$ & Unexpected \\
Themis & 2202 & C & 0.067 & 0.077 & $-0.009$ & Unexpected \\
\midrule
\multicolumn{7}{c}{\textit{Bimodal Families (likely mixing artifact)}} \\
\midrule
Nysa-Polana & 3184 & S+C & 0.221 & 0.100 & $+0.120$ & Mixing \\
Phocaea & 925 & S & 0.302 & 0.225 & $+0.077$ & Mixing \\
\bottomrule
\end{tabular*}
\begin{tablenotes}\small
\item Mean albedos $\langle p_V\rangle$ and $\Delta p_V$ are in linear 
$p_V$ (this analysis concerns absolute albedo differences, for which the 
linear scale is the natural one). All differences significant at 
$p < 10^{-3}$. Positive $\Delta p_V$ (small brighter) is consistent with 
space weathering; 
negative $\Delta p_V$ (large brighter) is unexpected and may indicate 
selection bias or unknown systematic effects. For bimodal families, 
the difference more likely reflects population mixing than physical 
processes. \textbf{Note:} Independent verification required.
\end{tablenotes}
\end{threeparttable}
\end{table*}

\subsection{Albedo Spreads: Observed vs Intrinsic}
\label{sec:albedo_spreads}
\label{sec:spreads}

\subsubsection{Comparison with Measurement Uncertainty}

\begin{figure}
\centering
\includegraphics[width=\columnwidth]{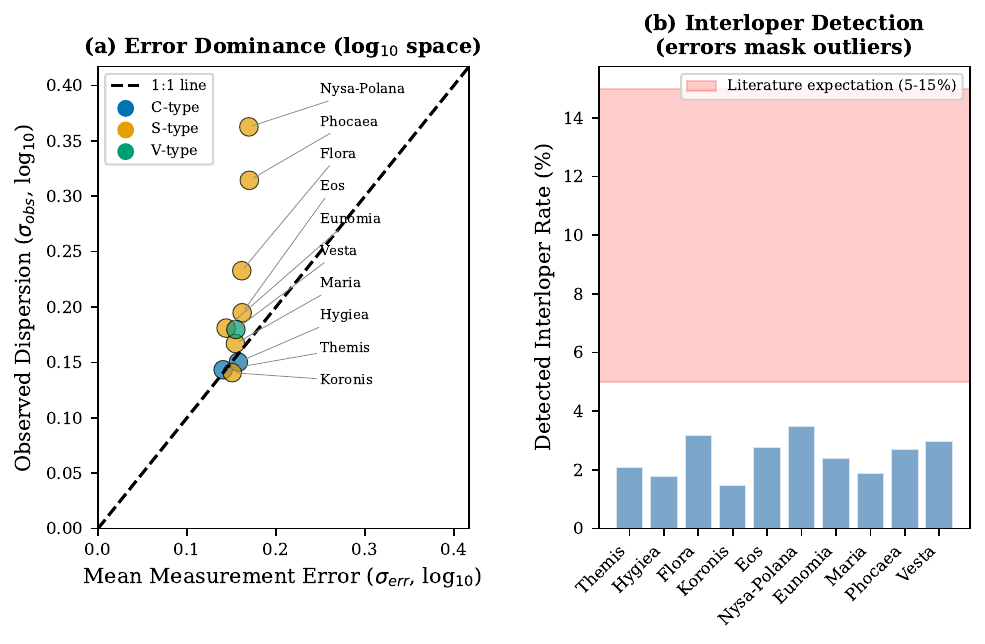}
\caption{NEOWISE error analysis in $\log_{10} p_V$. (a) Observed standard 
deviation ($\sigma_{\rm obs}$) vs mean measurement error ($\sigma_{\rm err}$) 
for nine major C- and S-complex families (the C- and S-type families of 
Table~\ref{tab:sigma} together with Nysa-Polana, Eunomia, Maria, and Phocaea); 
dashed line is 1:1. 
(b) Interloper detection rates from $\sigma$-clipping (blue bars); shaded 
band is the literature-expected range (5--15\%).}
\label{fig:error}
\end{figure}

Across the 102 families the median observed-to-error dispersion ratio in
$\log_{10} p_V$ is $\sigma_{\rm obs}/\sigma_{\rm err} \approx 0.99$ (mean 1.10;
Table~\ref{tab:sigma}, Figure~\ref{fig:error}). For 52\% of families
(53/102) the measurement error alone equals or exceeds the observed spread. Within-family
albedo scatter is therefore dominated by measurement error rather than intrinsic
compositional diversity. This holds in both linear
($\approx 1.03$) and logarithmic ($\approx 0.99$) representations, so it does not 
depend on the choice of albedo scale. The error dominance also suppresses
interloper detection. Our $\sigma$-clipping flags only 1--4\% of members as
outliers, well below the 5--15\% expected from the literature
(Figure~\ref{fig:error}b), because measurement errors inflate within-family
scatter and hide true compositional outliers.

\begin{table}[!ht]
\caption{Observed vs Intrinsic Spreads}
\label{tab:sigma}
\centering
\begin{threeparttable}
\begin{tabular*}{\tblwidth}{@{}LCRRRR@{}}
\toprule
Family & Type & \sigobs & \sigerr & \sigint & \sigint/\sigobs \\
\midrule
Themis & C & 0.143 & 0.141 & 0.028 & 0.20 \\
Hygiea & C & 0.150 & 0.158 & 0.000 & 0.00 \\
Flora & S & 0.233 & 0.162 & 0.168 & 0.72 \\
Koronis & S & 0.141 & 0.151 & 0.000 & 0.00 \\
Vesta & V & 0.180 & 0.155 & 0.091 & 0.51 \\
Eos & S & 0.195 & 0.162 & 0.108 & 0.56 \\
\bottomrule
\end{tabular*}
\begin{tablenotes}\small
\item Spreads in $\log_{10} p_V$. \sigerr{} is the RMS of the per-object 
propagated log error $0.434\,\sigma_{p_V,i}/p_{V,i}$; \sigint{} $= 
\sqrt{\max(0,\sigma^2_{\rm obs} - \sigma^2_{\rm err})}$. Across all 102 
families the median \sigobs/\sigerr{} $= 0.99$ (mean 1.10; 53/102 families 
have \sigobs{} $\leq$ \sigerr{}). Hygiea and Koronis have \sigobs{} 
$\lesssim$ \sigerr{}, yielding \sigint{} $\approx 0$.
\end{tablenotes}
\end{threeparttable}
\end{table}

We therefore report observed spreads (\sigobs) rather than attempting to 
infer intrinsic compositional variance. Comparisons between families should 
be interpreted in terms of measured albedo distributions rather than 
as estimates of intrinsic compositional spread.

\subsubsection{C-complex vs S-complex}

\begin{figure}
\centering
\includegraphics[width=\columnwidth]{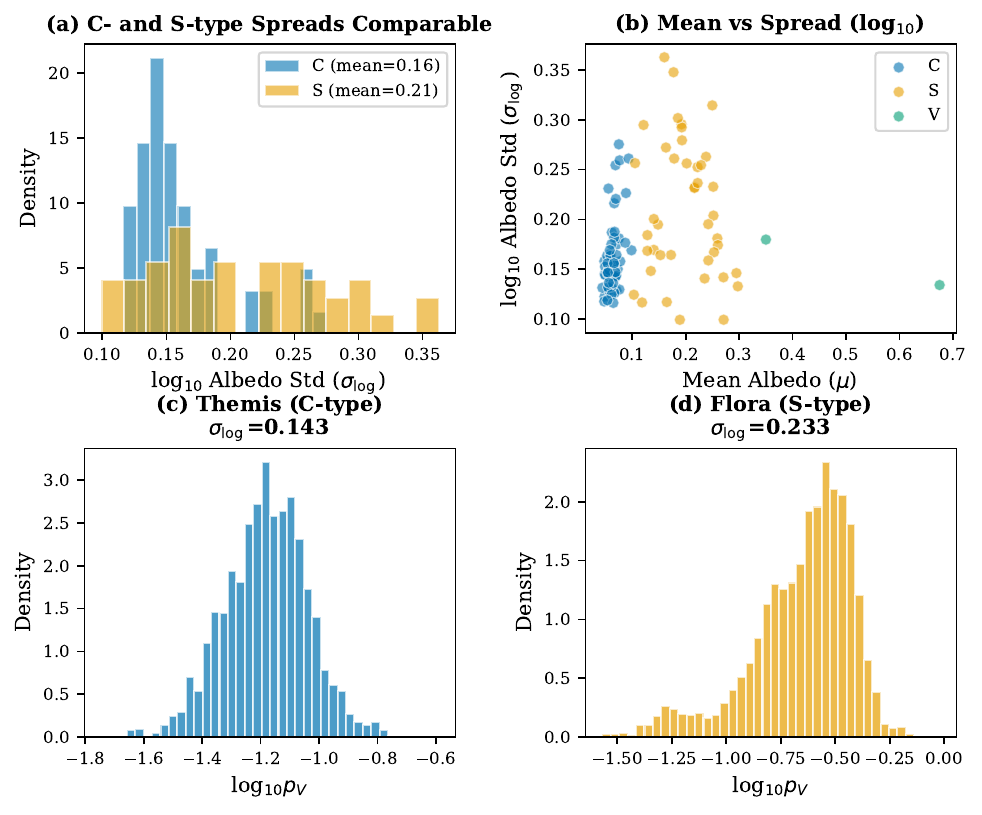}
\caption{Homogeneity comparison by spectral type in $\log_{10} p_V$. 
(a) Distribution of log-albedo standard deviations for C-complex (blue) and 
S-complex (orange) families; the means are comparable (0.16 vs 0.21). 
(b) Mean albedo vs.\ standard deviation per family. (c, d) Representative 
examples: Themis (C-complex) and Flora (S-complex).}
\label{fig:homogeneity}
\end{figure}

In $\log_{10} p_V$, C-complex and S-complex families show \emph{comparable} 
observed spreads. The mean log dispersion is $\sigma_{\rm obs} \approx 0.16$ 
(C-complex) versus $0.21$ (S-complex; Figure~\ref{fig:homogeneity}), a ratio of 
$\approx 1.3$. This is far smaller than the $\approx 3.6\times$ ratio in linear 
$p_V$ (median $\sigma_{\rm obs} = 0.024$ vs $0.086$). The large linear 
difference is therefore primarily a consequence of the two complexes' different 
mean albedos, not of intrinsically tighter C-complex distributions. A roughly 
constant fractional spread maps to a linear standard deviation proportional to 
the mean, so the low-albedo C-complex necessarily shows the smaller linear 
$\sigma_{\rm obs}$ (Section~\ref{sec:cv}). In log space, which removes this 
mean-albedo scaling, only a modest residual difference remains, with S-complex 
families slightly more dispersed. We therefore do not interpret the linear C/S 
contrast as evidence that C-complex families are intrinsically more homogeneous.

\subsubsection{Coefficient of Variation Interpretation}
\label{sec:cv}

The similar coefficients of variation (CV = $\sigma/\mu$, the ratio of 
standard deviation to mean) for C-complex 
and S-complex families (CV $\approx 0.4$) require careful interpretation. 
In an error-dominated regime where $\sigma_{\rm obs} \approx \sigma_{\rm err}$, 
the CV becomes approximately $\sigma_{\rm err}/\mu$. Since measurement 
errors are roughly constant but mean albedos vary considerably 
($\mu_C \approx 0.07$ vs $\mu_S \approx 0.23$), the similar CVs partly 
reflect this mathematical relationship rather than comparable intrinsic 
diversity.

Simulations confirm this effect. Populations with identical true CV 
($\sim$0.21) but different mean albedos show divergent observed CVs 
($\sim$0.55 for low-$\mu$ vs $\sim$0.27 for high-$\mu$) after adding 
realistic measurement errors.

The logarithmic analysis makes this explicit. In $\log_{10} p_V$, where the 
mean-albedo scaling is removed, the C- and S-complex spreads are comparable 
($0.16$ vs $0.21$; Section~\ref{sec:spreads}). The threefold difference in 
linear space is thus largely a coefficient-of-variation/mean-albedo effect 
rather than a difference in intrinsic compositional diversity.

\section{Validation}
\label{sec:validation}

\subsection{AKARI Cross-Validation}
\label{sec:akari}

\begin{figure}
\centering
\includegraphics[width=\columnwidth]{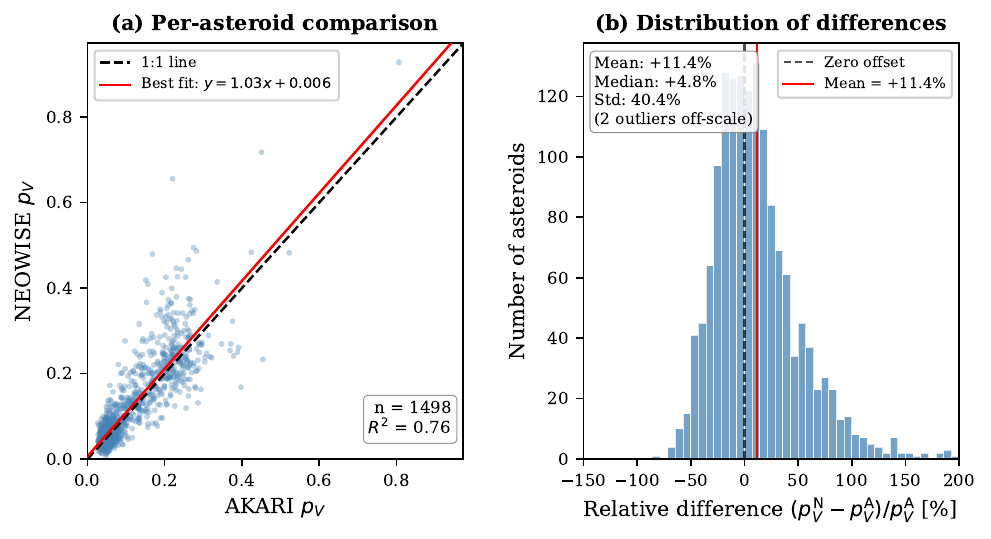}
\caption{NEOWISE versus AKARI visible albedo comparison for the 1{,}498 
matched asteroids. (a) Per-asteroid scatter with 1:1 line (dashed) and 
best-fit line (solid); fit parameters are reported in 
Section~\ref{sec:akari}. (b) Distribution of relative differences, 
$(p_V^{\rm NEOWISE} - p_V^{\rm AKARI})/p_V^{\rm AKARI}$; vertical line 
marks the mean offset.}
\label{fig:akari_comparison}
\end{figure}

\akari{} provides truly independent validation using different wavelengths 
and thermal models without relying on spectral classifications. We matched 
1{,}498 asteroids between catalogs (Figure~\ref{fig:akari_comparison}), of 
which 436 belong to Nesvorn\'y families, spanning 134 distinct 
families. Seven major families have sufficient AKARI sample ($n \geq 8$ 
matches per family) for the family-level cross-validation in 
Figure~\ref{fig:akari_validation}.

Direct per-asteroid comparison of the 1{,}498 matched objects 
(Figure~\ref{fig:akari_comparison}) shows broad agreement: the best-fit 
relation is $p_V^{\rm NEOWISE} = 1.03 \times p_V^{\rm AKARI} + 0.006$ with 
$R^2 = 0.76$. The median per-asteroid absolute difference is 0.015 (mean
0.027), and relative differences are centered at $+11.4\%$ (median $+4.8\%$) 
with standard deviation $40.4\%$. This scatter reflects measurement errors 
in both catalogs and the well-documented spectral-type-dependent offset. 
The offset is albedo dependent: low-albedo families (Nysa-Polana, Themis, 
Hygiea) agree to within $|\Delta| \leq 0.011$ in their mean albedos, whereas 
several high-albedo families show larger mean offsets up to $\sim 0.07$ 
(Table~\ref{tab:akari}), consistent with NEOWISE overestimating high albedos 
relative to AKARI \citep[the documented spectral-type dependence in 
NEOWISE thermal modeling;][]{Myhrvold2018b}. The near-unity best-fit slope confirms that NEOWISE and 
AKARI broadly agree on the same objects, with no large systematic skew that 
would create spurious albedo structure.

\subsubsection{Family-Level Distribution Agreement and Sampling Effects}

To assess cross-survey consistency at the family level, we compare AKARI 
and NEOWISE \pv{} distributions for individual families using the matched 
subsets (Figure~\ref{fig:akari_validation}). Restricting both surveys to 
the \textit{identical} asteroids is essential, because some AKARI 
family-level samples are not representative of the full NEOWISE families 
they belong to.

\begin{table}[!ht]
\caption{AKARI vs NEOWISE per-family agreement on matched asteroids}
\label{tab:akari}
\centering
\begin{threeparttable}
\begin{tabular*}{\tblwidth}{@{}LRRRRR@{}}
\toprule
Family & $n$ & $\langle p_V^{\rm N} \rangle$ & $\langle p_V^{\rm A} \rangle$ & Offset & Offset (\%) \\
\midrule
Nysa-Polana & 18 & 0.075 & 0.073 & $+0.002$ & $+2.7$ \\
Themis & 43 & 0.077 & 0.066 & $+0.011$ & $+16.7$ \\
Hygiea & 10 & 0.095 & 0.097 & $-0.002$ & $-2.1$ \\
Eos & 43 & 0.140 & 0.133 & $+0.007$ & $+5.3$ \\
Flora & 22 & 0.237 & 0.225 & $+0.012$ & $+5.3$ \\
Koronis & 6 & 0.235 & 0.220 & $+0.015$ & $+6.8$ \\
Eunomia & 9 & 0.179 & 0.139 & $+0.040$ & $+28.8$ \\
Phocaea & 15 & 0.226 & 0.171 & $+0.055$ & $+32.2$ \\
Maria & 5 & 0.264 & 0.198 & $+0.066$ & $+33.3$ \\
\midrule
All matched & 1498 & -- & -- & -- & $+11.4$ \\
\bottomrule
\end{tabular*}
\begin{tablenotes}\small
\item Mean NEOWISE ($\langle p_V^{\rm N} \rangle$) and AKARI 
($\langle p_V^{\rm A} \rangle$) albedos for the matched asteroids in each 
family, ordered by AKARI albedo. ``Offset'' is 
$\langle p_V^{\rm N} \rangle - \langle p_V^{\rm A} \rangle$. Agreement is 
good for low-albedo families ($|$offset$| \leq 0.011$) and larger for 
several high-albedo families, consistent with the known spectral-type 
dependence of the NEOWISE--AKARI offset. The small per-family samples 
($n = 5$--$43$) limit the precision of individual offsets; Eunomia, Phocaea, 
and Maria have both small samples and (for Eunomia) non-representative 
compositions, and should be interpreted with care.
\end{tablenotes}
\end{threeparttable}
\end{table}

Three of the nine families show significant AKARI sampling bias relative 
to their full NEOWISE membership: Nysa-Polana (AKARI matched subset 94\% 
low-albedo versus 55\% in the full family), Hygiea (30\% high-albedo in the 
matched subset versus 5\% in the full family), and Eunomia (33\% low-albedo 
versus 6\%). The remaining six families have AKARI subsets whose albedo 
composition matches the full family to within a few percentage points. The 
biased families are precisely those where naive family-level $\sigma$ 
comparisons between the two surveys are most misleading, since the AKARI 
subset samples a different part of the albedo distribution than the full 
family.

When the comparison is restricted to identical asteroids 
(Figure~\ref{fig:akari_validation}b,c), the AKARI and NEOWISE distributions 
overlap closely for both Nysa-Polana and Hygiea, confirming that the two 
surveys agree on the objects they jointly sample. The faint dashed reference 
curves showing the full NEOWISE family distributions make the AKARI sampling 
bias explicit: the matched-subset distributions deviate from the reference 
not because the surveys disagree on measurement values, but because the 
AKARI-matched asteroids are themselves a non-representative subset. 
Secondary effects---wavelength coverage (\akari{}'s 9--18~$\mu$m thermal 
bands versus \neowise{}'s 3.4--4.6~$\mu$m bands, which include reflected 
sunlight; \citealt{AliLagoa2018}) and differing beaming-parameter 
treatments---may contribute additional scatter but are subdominant to 
sample selection at the family level.

Hygiea's apparently wider AKARI spread reflects AKARI's preferential 
sampling of high-albedo objects within the family region: 30\% of the 
AKARI-matched Hygiea subset has $p_V > 0.12$, versus only 5\% in the full 
NEOWISE family. AKARI and NEOWISE measurements of these specific high-albedo 
objects agree closely (Figure~\ref{fig:akari_validation}c), supporting their 
interpretation as genuine interlopers rather than measurement artifacts.

\subsubsection{Confirmed Results}

\begin{figure*}
\centering
\includegraphics[width=\textwidth]{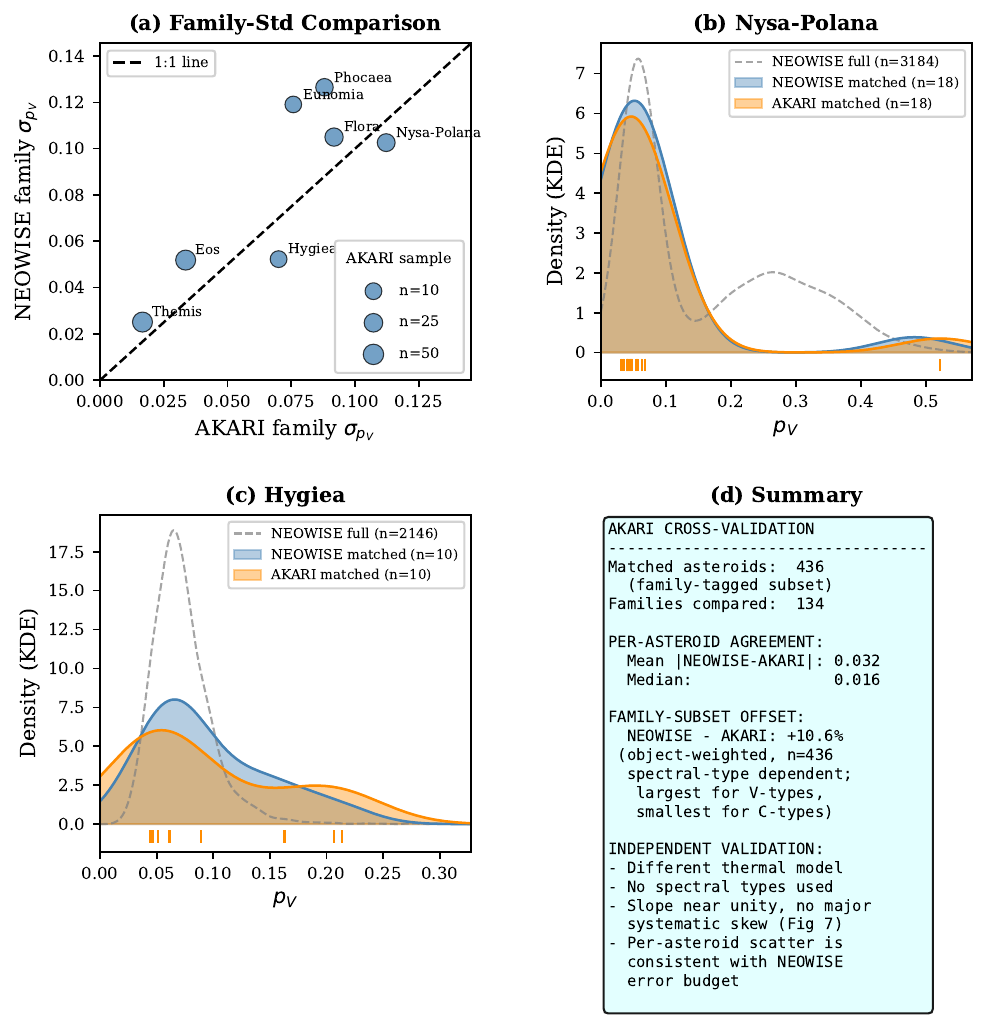}
\caption{AKARI--NEOWISE family-level cross-validation. 
(a) Within-family albedo standard deviations: AKARI vs NEOWISE, for the 
seven families with $n \geq 8$ AKARI matches. Marker area scales with AKARI 
sample size. (b) Nysa-Polana albedo distributions: full NEOWISE family 
(dashed gray, $n = 3{,}184$), NEOWISE restricted to the AKARI-matched 
subset (blue, $n = 18$), and AKARI for the same asteroids (orange). Rug 
ticks below the axis mark individual AKARI measurements. (c) Same overlay 
for Hygiea (matched $n = 10$). (d) Summary statistics for the 
family-tagged matched subset (Section~\ref{sec:akari}).}
\label{fig:akari_validation}
\end{figure*}

\akari{} cross-validation supports our conclusions in three ways. First, 
broad cross-survey agreement on the 1{,}498 matched asteroids (slope 1.03, 
$R^2 = 0.76$, median $|\Delta p_V| = 0.015$; Figure~\ref{fig:akari_comparison}) 
confirms that NEOWISE \pv{} measurements are not subject to large systematic 
skews that would create false bimodal structure. Second, AKARI independently 
supports the Nysa-Polana bimodal classification ($b = 0.99$ in AKARI data), 
validating our primary detection. Third, the apparent bimodal structure in 
the small AKARI Hygiea subset ($n = 10$; Figure~\ref{fig:akari_validation}c) 
reflects AKARI's preferential matching of high-albedo objects in the Hygiea 
family region rather than a true bimodal Hygiea family. AKARI and NEOWISE 
agree on these specific objects, supporting their interpretation as 
interlopers.

The two surveys agree on Nysa-Polana as bimodal, providing independent 
confirmation of our primary detection. Juno's bimodality is not recovered in 
the AKARI data, but the AKARI Juno sample is too small ($n < 5$ matches) to 
test for the asymmetric secondary population identified in the NEOWISE data. 
The absence of detection in AKARI thus reflects statistical power, not 
disagreement. More generally, asymmetric bimodality is difficult to detect 
reliably with current data, and results near detection thresholds depend on 
survey-specific sample sizes and error properties.

\subsection{SDSS Photometric Consistency}
\label{sec:sdss}

\subsubsection{The Circularity Problem}

Both \neowise{} albedos and SDSS-based photometric classifications 
\citep{Sergeyev2021} reflect surface composition, creating potential 
circularity in cross-validation. Classification confidence correlates 
with measured albedo ($\rho = 0.42$, $p < 10^{-15}$), indicating these 
are not fully independent.

We therefore interpret \sdss{}-\neowise{} comparisons as consistency 
checks rather than independent validations.

\subsubsection{Taxonomic Validation of Bimodality}

To validate that albedo bimodality reflects compositional mixing rather 
than measurement artifacts, we examined SDSS photometric classifications 
for bimodal and unimodal families (Table~\ref{tab:spectral_validation}). 
The ``mixing index'' (100\% minus dominant type fraction) quantifies 
taxonomic diversity.

\begin{table}[!ht]
\caption{Taxonomic Composition of Bimodal vs Unimodal Families}
\label{tab:spectral_validation}
\centering
\begin{threeparttable}
\begin{tabular*}{\tblwidth}{@{}LRRRRCR@{}}
\toprule
Family & $n_{\rm spec}$ & S\% & C\% & X\% & Dom & Mixing \\
\midrule
\multicolumn{7}{c}{\textit{Genuinely Bimodal (log space)}} \\
Nysa-Polana & 1321 & 35.7 & 26.8 & 2.3 & S & 64.3\% \\
Juno & 171 & 32.7 & 26.9 & 7.0 & S & 67.3\% \\
Phocaea & 325 & 56.9 & 11.4 & 4.6 & S & 43.1\% \\
Tirela & 168 & 17.3 & 7.1 & 7.1 & D & 70.2\% \\
Henan & 76 & 53.9 & 9.2 & 10.5 & S & 46.1\% \\
Telramund & 43 & 65.1 & 2.3 & 11.6 & S & 34.9\% \\
\midrule
\multicolumn{7}{c}{\textit{Unimodal S-type}} \\
Koronis & 474 & 81.0 & 0.8 & 2.7 & S & 19.0\% \\
Eunomia & 1061 & 75.1 & 2.1 & 2.9 & S & 24.9\% \\
Flora & 1426 & 57.6 & 8.0 & 13.7 & S & 42.4\% \\
Maria & 417 & 67.6 & 2.4 & 2.9 & S & 32.4\% \\
Hansa & 145 & 76.6 & 0.7 & 1.4 & S & 23.4\% \\
\midrule
\multicolumn{7}{c}{\textit{Unimodal C-type}} \\
Themis & 940 & 1.5 & 49.4 & 8.8 & C & 50.6\% \\
Hygiea & 957 & 1.4 & 47.1 & 8.9 & C & 52.9\% \\
Veritas & 377 & 3.2 & 49.1 & 24.7 & C & 50.9\% \\
Adeona & 713 & 5.3 & 55.7 & 9.7 & C & 44.3\% \\
\bottomrule
\end{tabular*}
\begin{tablenotes}\small
\item Mixing index = 100\% $-$ dominant type fraction (Dom = dominant 
spectral complex; X\% includes X-complex and other minor types not shown 
individually). Higher values indicate more taxonomic diversity. Sample sizes 
($n_{\rm spec}$) reflect asteroids with assigned SDSS-photometric 
classifications. For families also listed in 
Appendix Table~\ref{tab:spectral_fractions}, counts may differ by up to 
$\sim$1\%, as that table requires both the classification and its associated 
probability to be defined. Mean mixing indices quoted in the text are averaged over 
all classified families in each category, not only the representative 
families listed here.
\end{tablenotes}
\end{threeparttable}
\end{table}

The six genuinely bimodal families show substantially higher spectral mixing 
(mean 54.3\%) than either unimodal S-type (28.4\%) or unimodal C-type (49.7\%) 
families. Nysa-Polana contains both S-type (35.7\%) and C-type (26.8\%) 
objects, confirming that its albedo bimodality corresponds to real taxonomic 
diversity. Juno, Phocaea, and Henan show a similar S/C mixture. Tirela is a 
notable exception: its dominant spectral type is D (primitive, low-albedo), 
not C, and S+C types together account for only 24\% of its classified 
members---suggesting its dark component may be a D-type instead of a C-type 
population, a compositional pathway distinct from the other five genuine 
families. This validates that our bimodality detections reflect genuine 
compositional structure rather than measurement artifacts, while showing 
that not all dark/bright mixtures share the same taxonomic origin.

\subsubsection{V-type Bias Decomposition}

The apparent V-type albedo deficit decomposes into two components when separated 
by the SDSS-photometric classification probability $p$ (from \citealt{Sergeyev2021}; 
higher values indicate more reliable V-type identification) shown in 
Table~\ref{tab:vtype}.

\begin{table}[!ht]
\caption{V-type Albedo by Classification Confidence}
\label{tab:vtype}
\centering
\begin{threeparttable}
\begin{tabular*}{\tblwidth}{@{}LRRR@{}}
\toprule
Confidence & $n$ & Mean \pv & Bias \\
\midrule
$p < 0.3$ & 3335 & 0.10 & $-75\%$ \\
$0.3 \leq p < 0.5$ & 276 & 0.24 & $-41\%$ \\
$0.5 \leq p < 0.7$ & 267 & 0.31 & $-23\%$ \\
$0.7 \leq p < 0.9$ & 323 & 0.34 & $-14\%$ \\
$p \geq 0.9$ & 418 & 0.35 & $-13\%$ \\
\midrule
Literature & -- & 0.40 & -- \\
\bottomrule
\end{tabular*}
\begin{tablenotes}\small
\item $p$ denotes the SDSS-photometric classification probability for V-type 
assignment from \citet{Sergeyev2021}. Bias is computed as 
(mean \pv\ $-$ 0.40)/0.40, with 0.40 the literature mean V-type albedo for 
the Vesta family.
\end{tablenotes}
\end{threeparttable}
\end{table}

\textbf{Misclassification contribution ($\sim$46\%):} Low-confidence 
``V-type'' classifications (which are likely S-type or X-type asteroids 
misidentified based on ambiguous colors) account for most of the apparent 
deficit.

\textbf{Thermal model contribution ($\sim$13\%):} High-confidence V-types 
still show a significant deficit, consistent with known limitations of NEATM 
for high-albedo objects \citep{Mainzer2011,Masiero2011}. During the NEOWISE 
Reactivation mission, the beaming parameter ($\eta$) must be held fixed at 
$0.95 \pm 0.2$ due to the availability of only a single thermal band (W2; 
\citealt{Masiero2021}). The model's assumption of constant beaming parameter 
and infrared-to-visible albedo ratio can systematically underestimate albedos 
for bright, basaltic surfaces whose thermal properties differ from the assumed 
value. Contributing factors include:
\begin{itemize}
    \item Inappropriate beaming parameters ($\eta$) for basaltic vs 
    silicate surfaces
    \item Different thermal inertia of basaltic materials
    \item Observation geometry effects from the concentration of V-types 
    in the high-inclination Vesta family
\end{itemize}

We report the V-type bias as ranging from $-13\%$ (thermal model only) 
to $-59\%$ (combined with misclassification).

\section{Discussion}
\label{sec:discussion}

\subsection{What Can We Reliably Conclude?}

Based on our analysis, we classify conclusions by robustness. All findings 
are limited to what we can infer from albedo distributions alone, without 
independent spectroscopic or compositional constraints.

\textbf{Robust findings} (independent of SDSS taxonomic classifications and 
supported by \akari{} or consistency checks):
\begin{itemize}
    \item Genuine bimodality is found in 5.9\% of families (6 of 102) in 
    log-albedo space, meeting all four criteria
    \item Two families (Nysa-Polana, Juno) are genuinely bimodal in both linear 
    and logarithmic albedo space and are our most secure detections; four more 
    (Phocaea, Tirela, Henan, Telramund) are resolved specifically in log space, 
    where the asymmetric dark/bright mixture is no longer underweighted by the 
    moment-based test (Section~\ref{sec:log10})
    \item Monte Carlo simulations in log space show that a dark/bright 
    bimodality at the characteristic separation ($\Delta\mu \approx 0.64$ dex) 
    is recovered with high probability for the intrinsic scatter present in the 
    families; detection collapses only for per-component scatter $\gtrsim 0.2$ 
    dex, so the low rate reflects compositional homogeneity rather than limited 
    sensitivity
    \item Measurement errors dominate observed spreads for most families (median 
    $\sigma_{\rm obs}/\sigma_{\rm err} \approx 0.99$ in log space)
    \item Apparent size-albedo correlations largely vanish under conservative 
    bias controls. Interpreting them as space weathering without independent 
    spectral confirmation remains speculative
    \item C-complex and S-complex families show comparable observed spreads in 
    log space ($0.16$ vs $0.21$); the larger difference in linear space is a 
    mean-albedo (coefficient-of-variation) effect
\end{itemize}

\textbf{Tentative findings} (dependent on SDSS taxonomic classifications; see Appendix~\ref{app:spectral} for reliability assessment):
\begin{itemize}
    \item S-type albedos match literature expectations ($+2\%$)
    \item V-types show systematic deficits ($-13\%$ to $-59\%$) from thermal 
    model limitations and classification uncertainties
\end{itemize}

\subsection{Methodological Limitations}
\label{sec:methodological_limitations}

Our analysis uses simple variance subtraction ($\sigma^2_{\rm int} = \sigma^2_{\rm obs} - \sigma^2_{\rm err}$) 
to estimate intrinsic spreads. This approach assumes equal measurement errors for all objects and 
does not account for potential correlations between albedo and measurement 
uncertainty (e.g., size-dependent error scaling). More sophisticated methods---hierarchical 
Bayesian modeling, size-dependent error treatment, or measurement error 
deconvolution---would provide better uncertainty quantification but require 
larger sample sizes than most families offer. For the error-dominated regime 
we document ($\sigma_{\rm obs}/\sigma_{\rm err} \approx 0.99$ in log space), these refinements 
would not change our primary conclusion: observed spreads largely reflect 
measurement scatter rather than intrinsic compositional diversity.

\subsection{Robustness to albedo representation}
\label{sec:log10}

We use $\log_{10} p_V$ as the primary representation throughout 
Section~\ref{sec:bimodality_detection}, and we compare with the linear-$p_V$ 
analysis to characterise how the choice of scale affects the results. Using 
identical criteria in both spaces, overall classification agreement is 22.5\% 
(23/102 families). The transform materially alters which families are flagged.

Two effects operate in opposite directions. The log transform removes the 
positive skew of linear albedo distributions, so high-Sarle families with no 
genuine second peak fall below threshold. The number satisfying any multi-peaked 
category drops from 41 in linear space to 21 in log. More importantly, the 
transform symmetrises genuine but asymmetric dark/bright mixtures, raising their 
Sarle coefficients above threshold, so four families become genuinely bimodal in 
log, namely Phocaea, Tirela, and Henan (hidden in linear space) together with 
Telramund (wide-unimodal in linear; Table~\ref{tab:log10_comparison}). Log space is thus 
simultaneously more selective against skewness-driven false positives and more 
sensitive to real asymmetric bimodality, yielding 6 genuine detections versus 2 
in linear space.

Nysa-Polana and Juno are genuine in both representations and are our most secure 
detections. Recovering only these two near-symmetric mixtures in linear space, 
versus six in log, is the empirical basis for using the logarithmic 
representation rather than treating the linear result as a conservative default. 
The choice is also physically motivated, since albedos are approximately 
log-normal/double-Rayleigh \citep{Wright2016} and the two-Gaussian mixture 
underlying our criteria and simulations is therefore better specified in log 
space. The three remaining linear-hidden families (Gefion, Rafita, Witt; 
Section~\ref{sec:bimodality_detection}, Table~\ref{tab:hidden}) illustrate the 
boundary case: elevated Sarle without a sufficiently resolved structural peak, 
so they remain weak bimodal rather than genuine in either representation.

\begin{table*}[!ht]
\caption{Bimodality Classification: Linear vs.\ Logarithmic Space}
\label{tab:log10_comparison}
\centering
\begin{threeparttable}
\begin{tabular*}{\tblwidth}{@{}LCCCCCC@{}}
\toprule
Family & $n$ & Linear & $\log_{10}$ & Sarle$_{\rm lin}$ & Sarle$_{\rm log}$ & Robust \\
\midrule
\multicolumn{7}{c}{\textit{Genuine in both representations}} \\
Nysa-Polana & 3184 & GENUINE & GENUINE & 0.70 & 0.69 & Yes \\
Juno & 355 & GENUINE & GENUINE & 0.58 & 0.70 & Yes \\
\midrule
\multicolumn{7}{c}{\textit{Resolved in log space (linear $\rightarrow$ GENUINE)}} \\
Phocaea & 925 & HIDDEN & GENUINE & 0.31 & 0.71 & --- \\
Tirela & 376 & HIDDEN & GENUINE & 0.45 & 0.70 & --- \\
Henan & 187 & HIDDEN & GENUINE & 0.53 & 0.71 & --- \\
Telramund & 104 & WIDE & GENUINE & 0.43 & 0.70 & --- \\
\midrule
\multicolumn{7}{c}{\textit{Linear-hidden, not promoted (remain weak in log)}} \\
Gefion & 748 & HIDDEN & WEAK & 0.34 & 0.74 & --- \\
Rafita & 283 & HIDDEN & WEAK & 0.38 & 0.67 & --- \\
Witt & 89 & HIDDEN & WEAK & 0.46 & 0.74 & --- \\
\bottomrule
\end{tabular*}
\begin{tablenotes}\small
\item Family-by-family classification underlying the comparison discussed 
above. ``Robust'' (Yes) indicates GENUINE classification in 
both representations. Massalia (unimodal in linear, weak in log) is omitted.
\end{tablenotes}
\end{threeparttable}
\end{table*}

\subsection{Additional Limitations}

Several additional limitations warrant mention:

\subsubsection{Heliocentric distance bias}

Our $D > 5$~km completeness threshold 
addresses magnitude-limited bias against small dark asteroids, but does 
not fully correct for distance-dependent effects. Dark asteroids at larger 
heliocentric distances may still be under-represented 
\citep[cf.][]{DeMeo2013}. However, the bias correction methodology of 
\citet{DeMeo2013} was developed for population-wide taxonomic distributions, 
applying corrections to 13,211 asteroids drawn from a sample of 34,503. 
Our analysis differs fundamentally. We examine within-family albedo 
distributions using 124,091 asteroids across 102 families, where members 
of each family share similar heliocentric distances by definition.

To test whether distance-dependent bias affects our conclusions, we compared 
bimodality rates (genuine plus weak) across Main Belt zones in $\log_{10} p_V$. 
The rate declines with heliocentric distance, from 35.3\% (6/17) in the inner 
belt ($a < 2.5$~AU) through 23.8\% (10/42) in the middle to 11.6\% (5/43) in the 
outer belt ($a > 2.82$~AU), but a Fisher's exact test on the inner-versus-outer 
comparison gives $p = 0.06$ (odds ratio 4.1), so the difference is not 
significant at the $\alpha = 0.05$ level. The six \emph{genuine} detections are, 
by contrast, distributed evenly across the three zones, with Nysa-Polana and 
Phocaea in the inner belt, Juno and Henan in the middle, and Tirela and 
Telramund in the outer. The residual gradient is therefore confined to the 
weak-bimodal families, predominantly skewness artifacts 
(Section~\ref{sec:results}), and does not affect the genuine detections that 
constitute our primary result. The lower outer-belt rate is also consistent 
with the darker, more C-complex-dominated composition of outer families (mean 
family $p_V \approx 0.09$ versus $\approx 0.20$ in the inner belt), which offers 
less dark/bright contrast for two-population structure, rather than with a 
detection bias against dark asteroids. Heliocentric distance bias therefore 
does not substantially affect our bimodality conclusions.

\subsubsection{Space weathering saturation}

Our null result for size-albedo 
correlations could partly reflect space weathering saturation in old 
families. If weathering effects saturate on timescales shorter than 
family ages, large asteroids would show no correlation even if weathering 
operates, because all surfaces---regardless of exposure time---reach the 
same equilibrium state. This interpretation does not conflict with our 
selection bias findings but represents an alternative physical explanation 
for the absence of detectable trends.

\subsection{Recommendations for Future Studies}
\label{sec:recommendations}

For \neowise{}-based family studies, we recommend:
\begin{enumerate}
    \item Report observed spreads. Do not claim compositional homogeneity.
    
    \item Test size-albedo correlations with diameter-limited subsamples 
    to catch selection bias
    
    \item Be aware of pipeline filtering effects: during NEOWISE Reactivation, 
    observations where W2 band flux contains $>10\%$ reflected sunlight are 
    excluded \citep{Masiero2021}. This filtering removes ~29\% of Main Belt 
    observations and may introduce bias in datasets containing high-albedo 
    asteroids (S-type, E-type).
    
    \item Analyze bimodal families with population-specific methods
    
    \item Report spectral-type biases as ranges reflecting classification 
    uncertainty
    
    \item Use \akari{} or independent data for validation when possible
    
    \item Interpret CV values cautiously in error-dominated regimes
    
    \item Treat candidate weathering/differentiation signals 
    (Tables~\ref{tab:weathering_candidates} and \ref{tab:differentiation}) 
    as hypotheses requiring independent confirmation, not established results
    
    \item Detecting more asymmetric compositional subpopulations 
    ($w_1{:}w_2 \ll 15{:}85$; Section~\ref{sec:detection_results}) requires 
    deeper or higher-precision surveys: \neowise{} cannot reliably resolve 
    such configurations within current error budgets, even when present
\end{enumerate}

\section{Conclusions}
\label{sec:conclusions}

Our analysis of 102 asteroid families reveals that \neowise{} albedo 
distributions are dominated by measurement uncertainties, requiring careful 
interpretation:

\begin{enumerate}
    \item Genuine bimodality is found in 6 of 102 families (5.9\%) in log-albedo 
    space. Nysa-Polana and Juno are bimodal in linear space as well, and thus 
    our most secure detections; Phocaea, Tirela, Henan, and Telramund are 
    asymmetric dark/bright mixtures resolved specifically in log space, where 
    their Sarle coefficients rise from $b = 0.31$--$0.53$ to $b \approx 0.70$. 
    Detection-efficiency simulations place these families in the high-detection 
    regime and indicate they are not error-induced artifacts. The scarcity of 
    bimodality therefore reflects genuine compositional homogeneity and not 
    limited sensitivity.
    
    \item Measurement errors dominate observed spreads (median 
    $\sigma_{\rm obs}/\sigma_{\rm err} \approx 0.99$ in log space), limiting 
    inference of intrinsic compositional variance; the conclusion is independent 
    of linear vs.\ logarithmic representation.
    
    \item Apparent size-albedo correlations largely reflect selection bias 
    rather than space weathering; only 2 of 63 tested families survive 
    conservative diameter-limited tests. Vesta is a notable candidate that 
    warrants follow-up with independent methods.
    
    \item Spectral-type albedo biases are significant: V-types show 
    $-13\%$ to $-59\%$ systematic deficit due to thermal model limitations 
    and classification uncertainties.
\end{enumerate}

These findings establish detection limits, bias controls, and error budgets 
for interpreting NEOWISE family albedos in error-dominated regimes.


\appendix

\section*{Appendix}

\section{Hartigan's Dip Test Comparison}
\label{app:hartigan}

To validate our choice of combining Sarle and KDE metrics, we tested Hartigan's 
dip test on several asymmetric two-peak families---Phocaea (genuinely bimodal 
in log space) and Gefion (weak bimodal in log; ``hidden'' in the linear-space 
cross-check). 
Hartigan's test evaluates unimodality by measuring the maximum difference 
between the empirical distribution and the best-fitting unimodal distribution.

\begin{table}[!ht]
\caption{Hartigan Dip Test Results}
\label{tab:hartigan}
\centering
\begin{threeparttable}
\begin{tabular*}{\tblwidth}{@{}LRRL@{}}
\toprule
Family & $n$ & Dip $p$-value & Our Classification \\
\midrule
Phocaea & 925 & 0.112 & Genuine (log) \\
Gefion & 748 & 0.092 & Weak (log) \\
Themis & 2202 & 0.166 & Unimodal \\
Eos & 3953 & 0.234 & Unimodal \\
Flora & 3364 & 0.198 & Unimodal \\
Eunomia & 2166 & 0.160 & Unimodal \\
Vesta & 1891 & 0.118 & Unimodal \\
\bottomrule
\end{tabular*}
\begin{tablenotes}\small
\item All $p$-values $>0.05$ indicate failure to reject unimodality. 
Hartigan's test does not detect the two-peak structure visible in KDE analysis 
for Phocaea and Gefion, likely because these families have asymmetric population 
weights (15:85) and moderate peak separation. This motivated our combination 
of moment-based and density-based metrics.
\end{tablenotes}
\end{threeparttable}
\end{table}

The failure of Hartigan's test to detect bimodality in families with clear 
two-peak KDE structure (Phocaea, Gefion; Table~\ref{tab:hartigan}) 
shows the benefit of combining moment-based and density-based metrics. Hartigan's test is optimized for detecting 
departures from strict unimodality but may miss asymmetric or low-contrast 
bimodal structures that are nonetheless statistically resolvable.

\section{KDE Parameter Sensitivity}
\label{app:kde}

To verify that our bimodality classifications are not artifacts of 
specific KDE parameter choices, we tested the sensitivity of valley 
prominence to bandwidth and smoothing variations. For three representative 
families (Nysa-Polana, Gefion, Koronis), we varied the bandwidth multiplier 
(0.7, 1.0, 1.3 $\times$ Scott's rule) and Gaussian smoothing parameter 
($\sigma = 2, 3, 4$ grid points).

Classifications are stable under reasonable parameter variations:
\begin{itemize}
    \item \textbf{Nysa-Polana} (genuine bimodal): Valley prominence 
    ranges from 0.94 to 0.95 across all parameter combinations, remaining 
    well above the 0.25 threshold; classification unchanged.
    \item \textbf{Gefion} (weak bimodal in log space; ``hidden'' in the 
    linear cross-check): Valley prominence ranges from 0.15 to 0.33 
    depending on parameters. While this crosses our 0.25 threshold in some 
    configurations, the family consistently shows two-peak structure in KDE 
    analysis. This illustrates that such borderline families occupy a 
    transitional regime where classification depends on parameter choices.
    \item \textbf{Koronis} (unimodal): Although valley prominence ranges 
    from 0.28 to 0.44, the secondary peak height remains below 20\% of 
    the primary peak across all configurations, maintaining its unimodal 
    classification.
\end{itemize}

These results show that our four-criterion approach (requiring 
Sarle $> 0.556$, valley $> 0.25$, secondary peak $> 20\%$, and 
separation $> 1.2\sigma$) provides more robust classifications than 
any single metric. Genuine bimodals like Nysa-Polana are stable across 
all parameters, while edge cases like Gefion appropriately receive a 
weak-bimodal designation reflecting their ambiguous status.

\section{Bootstrap Stability Analysis}
\label{app:bootstrap}
To test the robustness of bimodality classifications, we performed bootstrap 
resampling analysis (1,000 iterations per family). For each iteration, we 
resample the family's albedo measurements with replacement and apply the 
four-criterion classification. We report the dominant classification (most 
frequent across iterations) and stability percentage (fraction of iterations 
yielding the dominant class).

\begin{table*}[!ht]
\caption{Bootstrap Stability by Bootstrap-Dominant Classification}
\label{tab:bootstrap}
\centering
\begin{threeparttable}
\begin{tabular*}{\tblwidth}{@{}LCCCC@{}}
\toprule
Bootstrap-Dominant Class & $n$ Families & Mean Stability & Median Stability & Range \\
\midrule
Genuinely Bimodal & 7 & 86.1\% & 97.6\% & 100.0--55.5\% \\
Weak Bimodal & 15 & 72.6\% & 70.6\% & 99.6--43.3\% \\
Wide Unimodal & 8 & 56.0\% & 55.4\% & 95.2--24.9\% \\
Skewed Unimodal & 26 & 75.8\% & 75.6\% & 100.0--33.0\% \\
Unimodal & 46 & 80.1\% & 88.2\% & 100.0--39.5\% \\
\midrule
All Families & 102 & 76.4\% & 77.1\% & 100.0--24.9\% \\
\bottomrule
\end{tabular*}
\begin{tablenotes}\small
\item Bootstrap stability quantifies classification robustness under 
resampling (1{,}000 iterations per family) in $\log_{10} p_V$, using the same 
five-class scheme as the point classification (Table~\ref{tab:bimodality}); no 
residual ``hidden bimodal'' category arises. The ``Bootstrap-Dominant Class'' 
is the most frequent classification across iterations and may differ from the 
point classification for borderline families (seven bootstrap-dominant 
genuine families versus six in Table~\ref{tab:bimodality}; the additional 
family, Juliana ($n=43$), has a bootstrap-dominant classification of 
Genuinely Bimodal at only 55.5\% stability, reflecting its small sample size 
and marginal position near the classification boundary). In total, 42 
families (41.2\%) fall below 70\% stability. Wide unimodal families show the 
lowest median stability (55.4\%), including the single least-stable family in 
the full sample (24.9\%), reflecting their ambiguous position near the 
bimodality threshold.
\end{tablenotes}
\end{threeparttable}
\end{table*}

These results (Table~\ref{tab:bootstrap}) demonstrate that: (1) the six 
genuine families are highly robust, with mean stability 91\% (individually 
67--100\%; the lowest, Telramund, reflects its small size, $n=104$), (2) the 
families that are weak bimodal in log space but were ``hidden'' in linear 
space---Gefion, Rafita, and Witt---remain stable ($>88\%$), (3) classification 
uncertainty is concentrated in families with weak signals or ambiguous 
distributions (wide unimodal, some skewed unimodal), and (4) the 
four-criterion approach successfully identifies clear cases while flagging 
borderline families through lower stability.

\section{Taxonomic Classification Reliability}
\label{app:spectral}

We evaluated the reliability of SDSS-based photometric classifications by 
comparing binary (counting) versus probability-weighted fractions for 
each family (Table~\ref{tab:spectral_fractions}). The weighted fraction 
accounts for classification uncertainty by weighting each asteroid's 
contribution by its classification probability.

Systematic differences between methods reveal classification reliability:
\begin{itemize}
    \item C-complex: Mean difference $-0.22$, indicating substantial 
    classification uncertainty (effective reliability $\sim$34\%)
    \item S-complex: Mean difference $-0.10$, moderate reliability ($\sim$68\%)
    \item V-type: Mean difference $-0.06$, higher reliability ($\sim$75\%)
\end{itemize}

Dominant taxonomic type agrees between methods for 88\% of families 
(60/68 families having sufficient SDSS taxonomic coverage for this 
comparison). Discrepancies occur primarily in families with ambiguous 
or mixed compositions.

\begin{table*}[!ht]
\caption{Probability-Weighted Taxonomic Fractions for Major Families}
\label{tab:spectral_fractions}
\centering
\begin{threeparttable}
\begin{tabular*}{\tblwidth}{@{}LCCCCCCC@{}}
\toprule
Family & $n$ & Dom & $f_S^{\rm bin}$ & $f_S^{\rm wgt}$ & $f_C^{\rm bin}$ & $f_C^{\rm wgt}$ & Note \\
\midrule
Koronis & 473 & S & 0.81 & 0.55 & 0.01 & 0.00 & Pure S \\
Eunomia & 1050 & S & 0.76 & 0.54 & 0.02 & 0.00 & Pure S \\
Gefion & 367 & S & 0.76 & 0.52 & 0.04 & 0.01 & S-dom \\
Vesta & 795 & V & 0.20 & 0.13 & 0.02 & 0.00 & Pure V \\
Flora & 1416 & S & 0.58 & 0.41 & 0.08 & 0.02 & S-dom \\
Themis & 937 & C & 0.01 & 0.00 & 0.50 & 0.17 & Pure C \\
Hygiea & 954 & C & 0.01 & 0.00 & 0.47 & 0.14 & Pure C \\
Nysa-Polana & 1314 & S & 0.36 & 0.23 & 0.27 & 0.08 & Bimodal \\
Eos & 1956 & S & 0.36 & 0.15 & 0.06 & 0.01 & K-type$^a$ \\
\bottomrule
\end{tabular*}
\begin{tablenotes}\small
\item $f^{\rm bin}$ = binary counting fraction; $f^{\rm wgt}$ = 
probability-weighted fraction. $^a$Eos is known as K-type but appears 
S-dominated in SDSS photometric classifications, likely reflecting spectral 
overlap between K and S types in broadband photometry. Sample sizes are 
restricted to asteroids with both a classification and a defined classification 
probability, and are therefore slightly smaller than the corresponding counts 
in Table~\ref{tab:spectral_validation}.
\end{tablenotes}
\end{threeparttable}
\end{table*}

Notable findings include:
\begin{enumerate}
    \item \textbf{Koronis} shows the highest S-type purity ($f_S^{\rm wgt} = 0.55$), 
    consistent with its reputation as a homogeneous S-type family.
    
    \item \textbf{Nysa-Polana} shows mixed composition ($f_S = 0.23$, $f_C = 0.08$), 
    consistent with its known bimodality.
    
    \item \textbf{Eos} appears S-dominant in SDSS but is known as K-type 
    from spectroscopy; this likely reflects spectral overlap between K and S 
    types in SDSS broadband photometry.
    
    \item \textbf{C-type families} (Themis, Hygiea) show large binary-weighted 
    differences, indicating that C-complex classifications have lower confidence 
    than S-complex.
\end{enumerate}

\section{Weak-Bimodal Families}
\label{app:weak}

Table~\ref{tab:weak} lists the fifteen families classified weak bimodal in
$\log_{10} p_V$ (Section~\ref{sec:weak}), together with their log-space
diagnostics and their classification in the linear cross-check. Only Gefion,
Massalia, and Chimaera show a KDE valley. In each the secondary peak is
below the 20\% height threshold, so none meets the genuine criteria. The remaining
twelve have no resolved second peak, and their high Sarle coefficients reflect
skewness rather than a distinct second component.

\begin{table*}[!ht]
\caption{Weak-Bimodal Families in Log Space}
\label{tab:weak}
\centering
\begin{threeparttable}
\begin{tabular*}{\tblwidth}{@{}LCCCCCL@{}}
\toprule
Family & $n$ & Sarle$_{\log}$ & Valley$_{\log}$ & 2nd/1st & Sep$_{\log}$ & Linear class \\
\midrule
Gefion & 748 & 0.74 & 0.91 & 0.15 & 2.72 & Hidden \\
Massalia & 293 & 0.69 & 0.86 & 0.18 & 2.59 & Unimodal \\
Rafita & 283 & 0.67 & -- & -- & -- & Hidden \\
Agnia & 128 & 0.67 & -- & -- & -- & Unimodal \\
Merxia & 115 & 0.65 & -- & -- & -- & Unimodal \\
Witt & 89 & 0.74 & -- & -- & -- & Hidden \\
Chimaera & 82 & 0.57 & 0.92 & 0.12 & 2.73 & Weak \\
Euterpe & 61 & 0.70 & -- & -- & -- & Skewed \\
Industria & 60 & 0.62 & -- & -- & -- & Unimodal \\
Marcello & 51 & 0.68 & -- & -- & -- & Unimodal \\
Gallia & 45 & 0.64 & -- & -- & -- & Unimodal \\
Juliana & 43 & 0.64 & -- & -- & -- & Weak \\
Tercidina & 34 & 0.57 & -- & -- & -- & Skewed \\
Patsy & 32 & 0.57 & -- & -- & -- & Unimodal \\
Helwerthia & 30 & 0.63 & -- & -- & -- & Weak \\
\bottomrule
\end{tabular*}
\begin{tablenotes}\small
\item Families classified weak bimodal in $\log_{10} p_V$ (Sarle $> 0.556$ but
failing at least one structural criterion). Columns: $n$ = members;
Sarle$_{\log}$, Valley$_{\log}$, 2nd/1st (ratio of secondary to primary KDE peak
height), and Sep$_{\log}$ (peak separation in units of the family $\log_{10} p_V$
standard deviation) are the log-space diagnostics; ``Linear class'' is the
classification of the same family in the linear cross-check
(Section~\ref{sec:log10}), abbreviated Hidden (hidden bimodal), Weak (weak
bimodal), Skewed (skewed unimodal), and Unimodal. A dash indicates that no second
KDE peak is resolved. Gefion, Massalia, and Chimaera have a KDE valley but a
secondary peak below the 20\% threshold, so none is promoted to genuine.
\end{tablenotes}
\end{threeparttable}
\end{table*}

\section*{Declaration of generative AI and AI-assisted technologies in the writing process} 

During the preparation of this work, the author used ChatGPT-4, Claude Opus
4.5, and Gemini AI to assist with language editing and readability improvements,
and to help generate Python code for stress tests and related analyses. After
using these tools/services, the author reviewed and edited the content as
needed and takes full responsibility for the content of the publication.


\section*{Declaration of competing interest}
The authors of the article declare they have no competing interests.

\printcredits

\section*{Acknowledgments}
This work uses data from NASA's NEOWISE mission, and JAXA's AKARI mission. 
Family classifications are from the \citet{Nesvorny2015} catalog and the 
2024 V2.0 update \citep{Nesvorny2024} via the Planetary Science Institute PDS Small Bodies Node. This research has made use of
the VizieR catalogue access tool, CDS, Strasbourg, France (DOI:
10.26093/cds/vizier). The original description of the VizieR service was
published in A\&AS 143, 23.

We thank the anonymous referees whose constructive comments substantially 
improved this work, including the contextualization of prior work in the 
Introduction, the formal multimode statistical validation of bimodality 
classifications, the analyses of detection efficiency and selection bias 
in size--albedo correlations, the treatment of measurement-error dominance, 
the heliocentric-distance and logarithmic-albedo robustness checks, and 
the V-type bias decomposition.

\paragraph{Software}
Python 3, NumPy \citep{numpy}, SciPy \citep{scipy}, 
Matplotlib \citep{matplotlib}, scikit-learn \citep{sklearn}, 
Pandas \citep{pandas}


\bibliographystyle{cas-model2-names}



\end{document}